\documentclass[twocolumn]{aastex701}
\usepackage{amsmath,graphicx,float,subcaption,multirow}
\usepackage{threeparttable}

\begin{document}


\title{
Evidence of self-organized criticality in the prompt emission 
of a bright gamma-ray burst
}

\correspondingauthor{Shuang-Xi Yi, Shao-Lin Xiong, R. Moradi, Fa-Yin Wang}
\author[0009-0008-6247-0645]{Wen-Long Zhang}
\affiliation{School of Physics and Physical Engineering, Qufu Normal University, Qufu, Shandong 273165, China}
\affiliation{State Key Laboratory of Particle Astrophysics, Institute of High Energy Physics, Chinese Academy of Sciences, Beijing 100049, China}
\affil{Purple Mountain Observatory, Chinese Academy of Sciences, Nanjing 210023, China}
\affil{School of Astronomy and Space Sciences, University of Science and Technology of China, Hefei 230026, China}
\email{wlzhang@pmo.ac.cn}

\author[0009-0006-5506-5970]{Wen-Jun Tan}
\affiliation{State Key Laboratory of Particle Astrophysics, Institute of High Energy Physics, Chinese Academy of Sciences, Beijing 100049, China}
\affiliation{University of Chinese Academy of Sciences, Chinese Academy of Sciences, Beijing 100049, China}
\affiliation{Department of Astronomy, School of Physics, Peking University, Beijing 100871, China}
\email{tanwj@ihep.ac.cn}

\author{Hao-Tian Lan}
\affiliation{School of Astronomy and Space Science, Nanjing University, Nanjing 210023, China}
\email{181840109@smail.nju.edu.cn}

\author[0000-0003-0672-5646]{Shuang-Xi Yi*}
\affiliation{School of Physics and Physical Engineering, Qufu Normal University, Qufu, Shandong 273165, China}
\email[show]{yisx2015@qfnu.edu.cn}

\author[0000-0002-4771-7653]{Shao-Lin Xiong*}
\affiliation{State Key Laboratory of Particle Astrophysics, Institute of High Energy Physics, Chinese Academy of Sciences, Beijing 100049, China}
\email[show]{xiongsl@ihep.ac.cn}

\author[0009-0008-8053-2985]{Chen-Wei Wang}
\affiliation{State Key Laboratory of Particle Astrophysics, Institute of High Energy Physics, Chinese Academy of Sciences, Beijing 100049, China}
\affiliation{University of Chinese Academy of Sciences, Chinese Academy of Sciences, Beijing 100049, China}
\email{cwwang@ihep.ac.cn}

\author[0000-0001-5586-1017]{Shuang-Nan Zhang}
\affiliation{State Key Laboratory of Particle Astrophysics, Institute of High Energy Physics, Chinese Academy of Sciences, Beijing 100049, China}
\affiliation{University of Chinese Academy of Sciences, Chinese Academy of Sciences, Beijing 100049, China}
\email{zhangsn@ihep.ac.cn}

\author{C. Guidorzi}
\affiliation{Department of Physics and Earth Science, University of Ferrara, Via Saragat 1, I-44122 Ferrara, Italy}
\email{guidorzi@fe.infn.it}

\author{R. Maccary}
\affiliation{Department of Physics and Earth Science, University of Ferrara, Via Saragat 1, I-44122 Ferrara, Italy}
\email{romain.maccary@edu.unife.it}

\author{R. Moradi*}
\affiliation{ICRANet Piazza della Repubblica 10 I-65122 Pescara Italy}
\affiliation{ICRA Dipartimento di Fisica Sapienza Universita di Roma Piazzale Aldo Moro 5 I-00185 Roma Italy}
\affiliation{State Key Laboratory of Particle Astrophysics, Institute of High Energy Physics, Chinese Academy of Sciences, Beijing 100049, China}
\email[show]{rmoradi@ihep.ac.cn}

\author[0000-0001-5798-4491]{Cheng-Kui Li}
\affiliation{State Key Laboratory of Particle Astrophysics, Institute of High Energy Physics, Chinese Academy of Sciences, Beijing 100049, China}
\email{lick@ihep.ac.cn}

\author[0000-0001-9217-7070]{Sheng-Lun Xie}
\affiliation{School of Mathematics and Physics, Jinggangshan University, Ji'an, Jiangxi 343009, China}
\affiliation{Institute for Astronomy and Astrophysics, Department of Physics, Jinggangshan University, Ji'an, Jiangxi 343009, China}
\email{xiesl@mails.ccnu.edu.cn}

\author[0000-0001-8664-5085]{Wang-Chen Xue}
\affiliation{State Key Laboratory of Particle Astrophysics, Institute of High Energy Physics, Chinese Academy of Sciences, Beijing 100049, China}
\affiliation{University of Chinese Academy of Sciences, Chinese Academy of Sciences, Beijing 100049, China}
\email{xuewc@ihep.ac.cn}

\author{Jia-Cong Liu}
\affiliation{State Key Laboratory of Particle Astrophysics, Institute of High Energy Physics, Chinese Academy of Sciences, Beijing 100049, China}
\affiliation{University of Chinese Academy of Sciences, Chinese Academy of Sciences, Beijing 100049, China}
\email{liujiacong@ihep.ac.cn}

\author{Zheng-Hang Yu}
\affiliation{State Key Laboratory of Particle Astrophysics, Institute of High Energy Physics, Chinese Academy of Sciences, Beijing 100049, China}
\affiliation{University of Chinese Academy of Sciences, Chinese Academy of Sciences, Beijing 100049, China}
\email{zhyu@ihep.ac.cn}

\author{Yue Wang}
\affiliation{State Key Laboratory of Particle Astrophysics, Institute of High Energy Physics, Chinese Academy of Sciences, Beijing 100049, China}
\affiliation{University of Chinese Academy of Sciences, Chinese Academy of Sciences, Beijing 100049, China}
\email{yuewang@ihep.ac.cn}

\author{Peng Zhang}
\affiliation{State Key Laboratory of Particle Astrophysics, Institute of High Energy Physics, Chinese Academy of Sciences, Beijing 100049, China}
\affiliation{Institute of Astrophysics, Central China Normal University, Wuhan 430079, China}
\email{zhangp97@ihep.ac.cn}

\author{Yan-Qiu Zhang}
\affiliation{State Key Laboratory of Particle Astrophysics, Institute of High Energy Physics, Chinese Academy of Sciences, Beijing 100049, China}
\affiliation{University of Chinese Academy of Sciences, Chinese Academy of Sciences, Beijing 100049, China}
\affiliation{School of Physics and Electronic Science, Guizhou Normal University, Guiyang 550001, People’s Republic of China}
\email{yqzhang@ihep.ac.cn}

\author{Chao Zheng}
\affiliation{State Key Laboratory of Particle Astrophysics, Institute of High Energy Physics, Chinese Academy of Sciences, Beijing 100049, China}
\affiliation{University of Chinese Academy of Sciences, Chinese Academy of Sciences, Beijing 100049, China}
\email{zhengchao97@ihep.ac.cn}

\author{Jin-Peng Zhang}
\affiliation{State Key Laboratory of Particle Astrophysics, Institute of High Energy Physics, Chinese Academy of Sciences, Beijing 100049, China}
\affiliation{University of Chinese Academy of Sciences, Chinese Academy of Sciences, Beijing 100049, China}
\email{zhangjinpeng@ihep.ac.cn}

\author[0000-0003-4157-7714]{Fa-Yin Wang*}
\affiliation{School of Astronomy and Space Science, Nanjing University, Nanjing 210023, China}
\email[show]{fayinwang@nju.edu.cn}

\begin{abstract}
Gamma-ray bursts (GRBs) are the most energetic explosive events in the Universe, yet the physical mechanism of their prompt emission remains a mystery. Especially, it is unclear whether the energy dissipation mechanism in the GRB jet is dominated by kinetic energy or magnetic energy. Here, we studied the pulses in the prompt emission of the second brightest GRB to date, GRB 230307A, which was accurately measured by the Gravitational wave high-energy electromagnetic counterpart all-sky monitor (GECAM), with focus on the cumulative distributions of peak counts and duration of pulses as well as the waiting time between pulses. We find that these cumulative distributions show scale-invariant behavior, well consistent with the prediction of the self-organized criticality (SOC) theory. This is the first robust evidence of an SOC feature in the prompt emission of a single GRB. Moreover, the statistical properties of pulses in the prompt emission of GRB 230307A are very similar to those of solar flares. Our findings suggest that the prompt emission of GRB is powered by the dissipation of magnetic energy in the ultra-relativistic jet, supporting the Poynting-flux-dominated prompt models.
\end{abstract}

\keywords{Gamma-ray bursts --- Self-organized criticality}


\section{INTRODUCTION}
\label{sec:intro}

The nature of the energy dissipation mechanism that powers the GRB prompt emission is still an intriguing mystery \citep{Piran2004,Zhang2014,Kumar2015,Daigne2025}. There are two competing models: The first is the traditional internal shock model \citep{Meszaros1993}. For a matter-dominated fireball, its expansion is driven by its own thermal pressure and accelerates to a relativistic speed, in which the thermal energy converts to the kinetic energy \citep{Shemi1990}. The kinetic energy of the outflow is dissipated by the collisions of shells, called internal shocks \citep{Rees1994,Sari1995}. For the second model, the GRB outflows are Poynting-flux-dominated \citep{Usov1992,Lyutikov2003,Giannios2005,Zhang2011,Metzger2011,Zhang2024}. Zhang \& Yan (2011) suggest that the initial shell collisions will distort the magnetic field configurations. As multiple collisions proceed, the magnetic field configuration would be distorted to a critical point at which a cascade of magnetic reconnection occurs \citep{2006Sci...311.1127D,Zhang2011}. Charged particles can be accelerated in these reconnection regions, producing prompt emission.

Some clues supporting the magnetic reconnection scenario have been found from statistical study. For example, similar statistical properties between GRB prompt emissions and solar X-ray flares indicate the common physical mechanism of them \citep{2013NatPh...9..465W,2021FrPhy..1614501L,2023ApJS..265...56L,2026ApJ..1006..236L,2024ApJ...965...72M}. And there is also statistical evidence for a common physical mechanism for GRB prompt emission and GRB X-ray flares \citep{2015ApJ...801...57G}. However, due to the lack of large sample of pulses from a single GRB, these works collected pulses from different GRBs, strongly affected by possible selection bias and systematic errors. Until now, whether an individual GRB prompt emission shows similar statistical properties has not yet been established.

GRB 230307A was first detected by the Gravitational-wave high-energy electromagnetic counterpart all-sky monitor (GECAM) at 15:44:06.650 UTC on March 7, 2023 (hereafter $T_0$), and was simultaneously confirmed by Fermi/GBM \citep{FermiGBMTeam07A_obs}. The GECAM team promptly issued a real-time alert reporting the exceptional brightness of this burst \citep{2023GCN.33406....1X}, which triggered extensive follow-up observations across multiple wavelengths. Thanks to its dedicated anti-saturation design, GECAM obtained an accurate measurement of the prompt emission without data loss, despite the extraordinarily high flux of the burst. Similar to GRB 211211A \citep{Rastinejad2022}, GRB 230307A was found to exhibit a kilonova signature \citep{2024Natur.626..737L,Yang2024}, indicating a compact binary merger origin. This classification is further supported by its identification as a long-duration Type I (Type IL) gamma-ray burst \citep{2024arXiv240702376W,2025ApJ...993...89T}. The unprecedented brightness of GRB~230307A has enabled detailed investigations of its temporal and spectral evolution during the prompt phase \citep{2023ApJ...953L...8W,2024ApJ...977..155M,2025NSRev..12E.401S}, and the high-quality light curve recorded by GECAM provides a unique opportunity to study pulse statistics within a single GRB. The multi-energy-band light curves reveal a distinct multi-pulse structure, and characterizing the statistical behavior of these pulses is essential for understanding the underlying physical processes of GRB prompt emission.

We identified pulses using MEPSA \citep{2015A&C....10...54G}, which is a powerful peak search algorithm for GRB light curves. For the eight energy bands (30-70, 70-100, 100-150, 150-200, 200-300, 300-500, 500-1000, 1000-6000 keV) of GECAM, each pulse is characterized by the pulse peak time $t_0$, peak counts, the signal to noise ratio ($S/N$), and the detection timescale. For the goal of the robust analysis, we selected the pulse candidates with $S/N>5$. After all pulses are identified, we derived the following parameters of pulses, including duration $T$ and peak counts $P$ of each pulse, and waiting time $\Delta t$ between adjacent pulses. The waiting time is calculated from $\Delta t=t_{\text{peak},i+1}-t_{\text{peak},i}$, where $t_{\text{peak},i+1}$ and $t_{\text{peak},i}$ are the peak times for the $(i + 1)$th and $i$th pulse. The detailed procedure to identify pulses and calculate their parameters are outlined in \textbf{Appendix}.

\section{OBSERVATIONS AND DATA ANALYSIS}
\label{sec:obs}

\begin{figure*}[htp!]
\centering
\includegraphics[width=0.8\textwidth, angle=0]{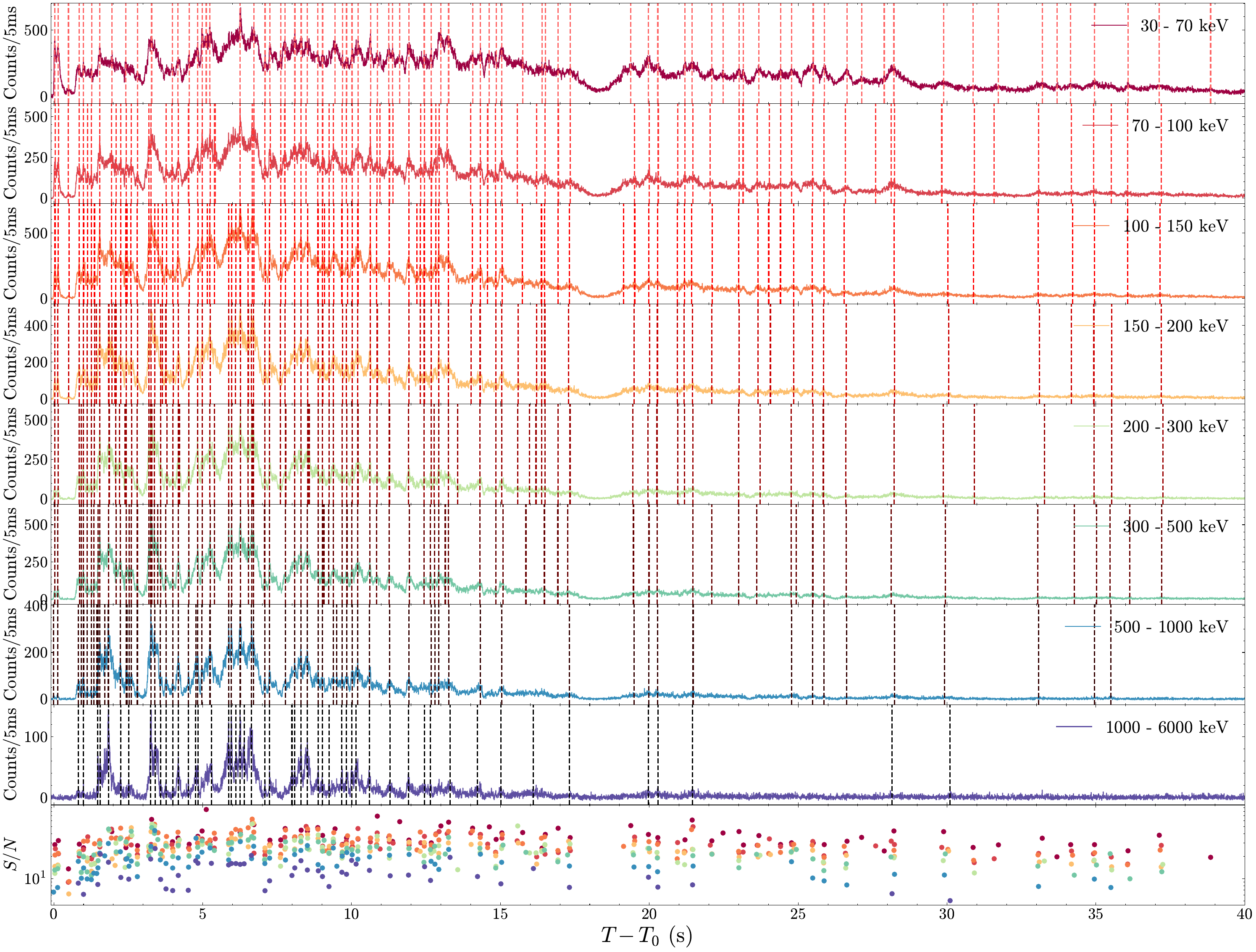}
\caption{\textbf{Peak identification results for GRB 230307A in various energy bands.} Each dashed line represents the peak time of each pulse. The signal to noise ratio $S/N$ of each peak is shown in the bottom panel.}
\label{LC_pks}
\end{figure*}

The peak identification result of the prompt emission of GRB 230307A is shown in Figure~\ref{LC_pks}. The $S/N$ of each peak is shown in the bottom panel of Figure~\ref{LC_pks}. The cumulative distribution of $S/N$ is shown in Extended Data Figure~\ref{SNR}. The number of pulses varies from 51 to 98 for different energy bands, as shown in Table~\ref{tab:fitting_results}. From this figure, one can see that many peaks concurrently appear across multiple energy bands. This feature is in line with that the spectral lag of most pulses is negligible \citep{2024arXiv240702376W}.

For small samples, the cumulative distribution analysis is often used, because the small number of pulses is not sufficient to bin the data. The cumulative distribution is defined by the integral of the total number of pulses above a given value $x_{\rm b}$. For a simple power-law function
\begin{equation}
N(x)dx=n_0 x^{-\alpha_x}dx,\quad x_1\leqslant x\leqslant x_2,
\label{diff_pl}
\end{equation}
where $x_1$ and $x_2$ are the lower and upper bounds of the power-law inertial range, $\alpha_x$ is the power-law slope, and $n_0=n_\mathrm{ev}\big(1-\alpha_x\big)/(x_2^{1-\alpha_x}-x_1^{1-\alpha_x})$ is the normalization constant with the event number $n_\mathrm{ev}$. With $N$ events in $x+dx$, the cumulative distribution is derived from the integral of Eq.~(\ref{diff_pl}),
\begin{equation}
N_{\mathrm{cum}} (>x) = n_0\times \left(x_{2}^{1-\alpha_x}-x^{1-\alpha_x}\right),
\label{cum_pl}
\end{equation}
where $n_0$, $x_2$ and $\alpha_x$ are the free parameters involved in the fitting process.

The cumulative distributions of waiting time $\Delta t$, duration $T$, and peak counts $P$ are shown in panels \textbf{a}, \textbf{b}, and \textbf{c} of Figure~\ref{Tpw_bands}, respectively. All distributions resemble a power-law function at high end. Eq. (2) is consistent with a power law function with slope (1-$\alpha_x$) and a size cutoff at $x<x_2$. However, the low end deviates from an ideal power-law function in form of a gradual rollover, which may be caused by incomplete sampling at the sensitivity limit \citep{2015ApJ...814...19A}. Eq.~(\ref{cum_pl}) is used to fit the data above $x_{\rm b}$ (due to incomplete sampling below $x_{\rm b}$). The value of $x_{\rm b}$ is determined by a bent power-law function with a natural break \citep{2016A&A...589A..98G}, which is widely used in distributions showing a break in the power-law form. It is equivalent to the power-law function in the limit $x\gg x_{\rm b}$. The best-fit curves are shown as solid lines in panels \textbf{a}, \textbf{b} and \textbf{c}, with the values of $\alpha_{x}$ given in Table~\ref{tab:fitting_results}. The fitting residuals are shown in Extended Data Figure~\ref{Residual}. The residuals are consistent with zero besides a few outliers. From panels \textbf{d} and \textbf{e} in Figure~\ref{Tpw_bands}, one can see that $\alpha_T$ and $\alpha_{\Delta t}$ have a large value in low-energy bands, and decrease to around $2.0$ in high-energy bands. The best-fit values of $\alpha_P$ of the peak counts in different energy bands are displayed in panel \textbf{f}, also showing a significant evolution across different energy bands. Its value decreases to around 1.7 in high-energy bands.

There is a natural explanation for the energy-dependence of best-fit power-law slopes ($\alpha_T$, $\alpha_{\Delta t}$ and $\alpha_P$) based on the distinctive pulse feature ``softer wider'' observed in GRB 230307A. More pulses are identified in the low-energy bands, which are also wider compared to those in the high-energy bands \citep{2025_mini_jet_Yi}. This can result in overlapping pulses and the inability of peak-finding algorithms to identify all peaks. However, as the energy increases, the pulses become narrower. Therefore, it is reasonable to attribute this evolution to the effect becoming negligible when the energy band surpasses a threshold value ($\sim$150 keV). If we consider the value of $x_{\rm b}$ of $\Delta t$ as a measure of pulse resolution, the pronounced decreasing trend of $x_{\rm b}$ with energy evolution in the high-energy range can serve as compelling evidence for this hypothesis. Similarly, the trend in the value of $x_{\rm b}$ of $T$ can also confirm that the pulses become narrower with increasing energy. Therefore, the results in high-energy bands can accurately reflect the actual parameter distributions, which will be discussed in detail in {\bf Appendix}.

\section{STATISTICAL ANALYSIS AND RESULTS}
\label{sec:results}

Here, we try to explain the cumulative distributions observed in the prompt emission of GRB 230307A using the self-organized criticality (SOC) theory \citep{Katz1986,1987PhRvL..59..381B,1988PhRvA..38..364B}. The concept of SOC was initially proposed to explain the prevalence of scale-free distributions in complex dynamical systems. We note, however, that a power-law distribution alone does not uniquely identify SOC, but only indicates the absence of a characteristic scale \citep{1941DoSSR..30..301K,Katz1986}. Nevertheless, as we show below, the observed distributions are not only consistent with power-law forms, but also their slopes match the quantitative predictions of 3D SOC models, providing stronger evidence beyond a simple scale-free behavior. SOC is a critical state of a complex system that is slowly and continuously driven toward a critical state, producing scale-free and intermittent avalanches showing power-law–like size distributions. Owing to some driving force, a SOC system will self-organize to a critical state at which a small perturbation can trigger an avalanche-like chain reaction. Approximating the original Bak-Tang-Wiesenfeld (BTW) model with a classical diffusion process \citep{2014ApJ...782...54A,2012A&A...539A...2A}, the statistical spatio-temporal scaling law between the size scale $L$ and the duration time $T$ of a SOC avalanche is \citep{2013NatPh...9..465W,2014ApJ...782...54A} 
\begin{equation}
L \propto T^{\beta/2},
\end{equation}
where $\beta$ is the spreading exponent, with $\beta=1$ for the classical diffusion. The probability distribution of size $L$ is found to be scale-free $N(L)dL \propto L^{-S}dL$ with the Euclidean dimensions $S$=1, 2 and 3 \citep{2014ApJ...782...54A}. Therefore, the power-law slope of the duration frequency distribution of pulses is 
\begin{equation}
\alpha_T ={1+(S-1)\beta/2}.
\label{th-T}
\end{equation}

The waiting time $\Delta t$ can reflect the occurrence rate of pulses and tell us whether these pulses occur independently or not. Recent studies have indicated that the waiting time distributions for solar flares and GRB X-ray flares exhibit a similar power-law-like behavior \citep{2013NatPh...9..465W,2016ApJS..224...20Y}, which was also discovered by \cite{2025A&A...693A.290Z} for TeV photons from GRB 221009A. In the temporal analysis of GRB 230307A, we define the waiting time ($\Delta t$) as the time interval between adjacent pulses. The distinct temporal separation of adjacent pulses in the light curve, combined with the continuous, uninterrupted nature of the pulse sequence, leads to a fundamental constraint in the tail region of the $\Delta t$ distribution (corresponding to statistically significant inter-pulse intervals): $\Delta t \geq T$. This inequality holds when the system operates in a steady-state continuous burst mode, characterized by the absence of quiescent phases and strict temporal isolation between consecutive events. Under these conditions, the waiting time distribution exhibits an isomorphism with the pulse duration distribution. So the power-law slope of the waiting time frequency distribution ($\alpha_{\Delta t}$) is \citep{2014ApJ...782...54A} 
\begin{equation}
\alpha_{\Delta t} =\alpha_T={1+(S-1)\beta/2}.
\end{equation}

For peak counts, according the statistical scaling law \citep{2014ApJ...782...54A} ${P \propto L^{d}}$, the power-law slope of the peak counts frequency distribution can be given by \citep{2014ApJ...782...54A}
\begin{equation}
\alpha_P =1+(S-1)/S.
\label{th-P}
\end{equation}
For classical diffusion $\beta=1$ in three dimensions $S=3$, $\alpha_{\Delta t}=\alpha_{T}=2$ and $\alpha_{P}=1.67$ are naturally expected \citep{2014ApJ...782...54A}. The values of $\alpha_{\Delta t}$, $\alpha_{T}$ and $\alpha_{P}$ for GRB 230307A in different energy bands are shown and plotted in Table~\ref{tab:fitting_results} and Figure~\ref{Tpw_bands} (panels {\bf d, e, f}), which are consistent with these theoretical predictions, as explained in {\bf Appendix} after considering the observed correlation (shown in Figure~\ref{Corr} {\bf a}) between duration and peak counts of all identified bursts.

It is well known that solar flares are energetic explosive phenomena in the solar atmosphere, which are widely believed to be triggered by magnetic reconnection \citep{2011LRSP....8....6S}. Statistical studies have shown that the duration and peak fluxes frequency distributions of solar flares are power-law functions with similar slopes as gamma-ray pulses detected in GRB 230307A \citep{2013NatPh...9..465W,2014ApJ...782...54A}. These scale-invariant distributions can be well explained by a three-dimension SOC process, which is confirmed by analytical and numerical models \citep{2013NatPh...9..465W,Harko2015,Danila2015}. The physical picture is that solar fares are avalanches of many small reconnection events. Flares of all sizes are manifestations of the same physical processes, where the size of a large flare is determined by the number of elementary reconnection events. Similar process can occur in the prompt emission of GRBs. For magnetic reconnection, the pulse energy will be inversely proportional to its characteristic time, which is the layer thickness at the moment of reconnection divided by the Alfvén velocity ($c$). Therefore, $E \propto T^{-1}$ is expected from theory. Our data also supports this expectation. From Figure~\ref{Corr}, the correlation $T\propto P^{-1/2}$ is found in lower energy bands. The pulse energy can be roughly estimated as $E\sim P T\propto T^{-1}$. For high-energy bands, the index slightly deviates from $-1$, which is consistent with $-1$ in $2\sigma$ confidence level.

In the internal shock model, a series of collisions can occur between the late, fast shells and the early, slow shells. These collisions will generate internal shocks from which particles are accelerated and photons are released to power the GRB prompt emission \citep{Rees1994}. Internal shocks can produce the observed variable temporal structure of the bursts from numerical simulations \citep{Kobayashi1997,Daigne1998,Nakar2002}. Moreover, for the equal-energy model, the correlation between pulse width $T$ and waiting time $\Delta t$ can be well produced \citep{Nakar2002}. However, if the relativistic shells collide randomly, we found that the waiting time distribution will be of exponential form by numerical simulations (see \textbf{Appendix}), which conflicts with the observed one. This scenario also can be derived theoretically \citep{Wheatland1998}. If the energy dissipation process follows a non-stationary Poisson process, the distribution of waiting times shows a power-law form, which is supported by cellular automaton simulations (see \textbf{Appendix}). Therefore, the consecutive pulse arriving times following a Poisson process cannot reproduce the observed waiting time distribution \citep{2015ApJ...801...57G}.

Similar frequency distributions of duration, peak counts and waiting time have been found between the pulses of GRB prompt emission and solar flares, which favor magnetic energy dissipation as GRB prompt emission trigger mechanism, such as magnetic reconnection. Theoretically, it has been long speculated that the GRB prompt emission is powered by magnetic energy \citep{Thompson1994,Lyutikov2003,Metzger2011,Zhang2011}. Here, we focus on the internal-collision-induced magnetic reconnection and turbulence (ICMART) model \citep{Zhang2011}, in which the ejecta launched from the GRB central engine is highly magnetized. The central engine (a black-hole–torus system or a rapidly spinning neutron star) is intermittent, launching an unsteady ejecta with variable luminosity and Lorentz factor, colliding with each other. These collisions distort the ordered magnetic field lines, and create small-scale regions of radiation, known as ``mini-emitters'', resemble the elementary reconnection events in solar flares \citep{1991ApJ...380L..89L}. These mini-emitters arise from magnetic instabilities, such as tearing modes and plasmoid formation, which occur as magnetic field lines reconnect \citep{2013arXiv1302.3254L}. At large radius, the critical condition to trigger a runaway magnetic reconnection avalanche is satisfied, leading to the bright pulse in the prompt emission. The turbulence within the jet facilitates the isotropic, three-dimensional expansion of these emitters, resulting in a system that exhibits SOC dynamics, characterized by a critical dimension of $S=3$. There are some other GRB prompt emission models, including electromagnetic model \citep{Lyutikov2003,Lyutikov2006} and dissipative sub-photospheric model \citep{Spruit2001,Drenkhahn2002,Giannios2008}. The jets are also Poynting-flux-dominated in these models. Although they face some challenges to explain the observed GRB properties \citep{Zhang2011}, the dissipation of magnetic energy to power GRB prompt emission is supported by this work.

\begin{figure*}[htp]
\centering
\includegraphics[width=0.9\textwidth, angle=0]{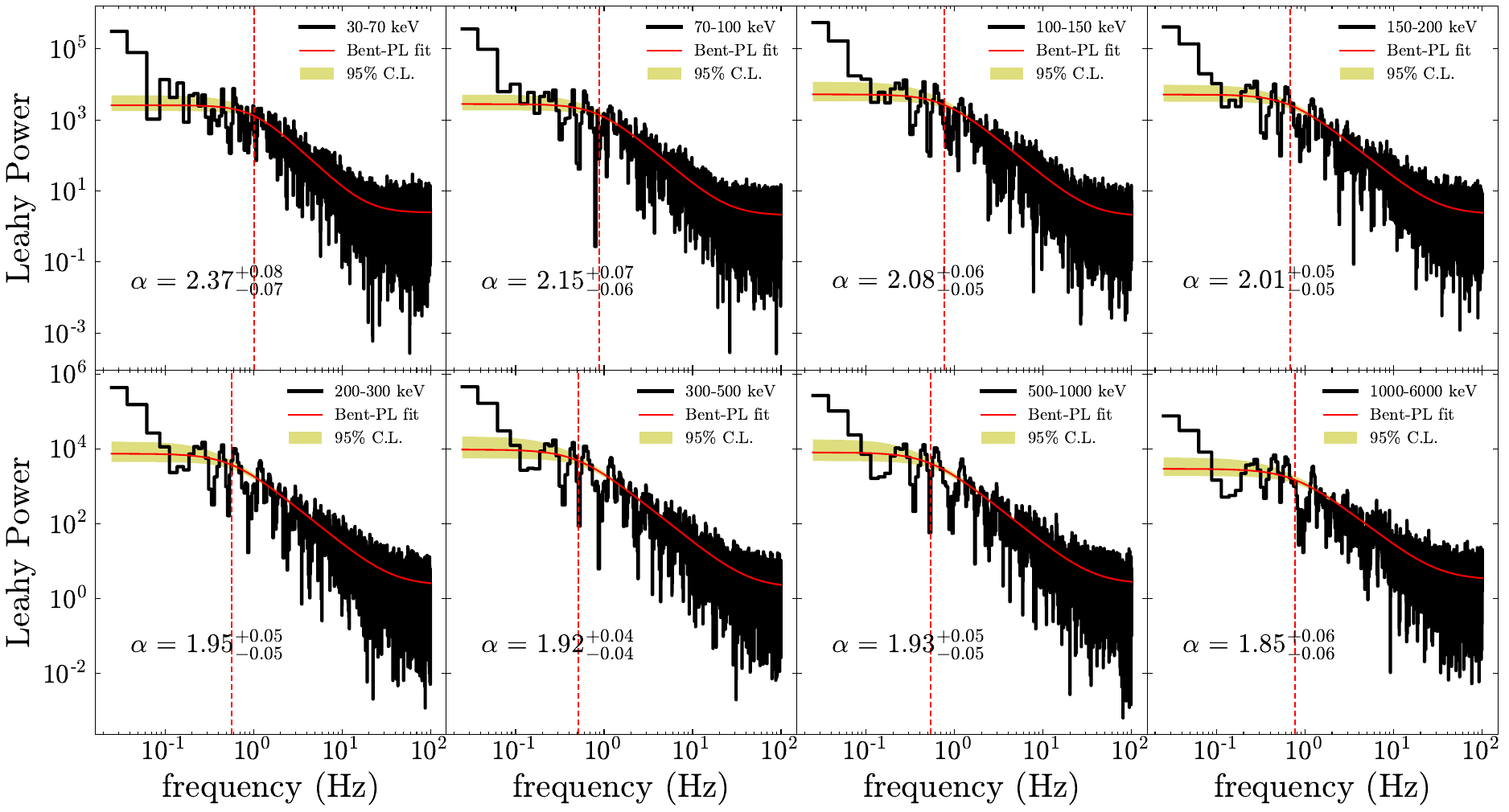}\put(-450, 250){\bf a}\\
\includegraphics[width=0.9\textwidth, angle=0]{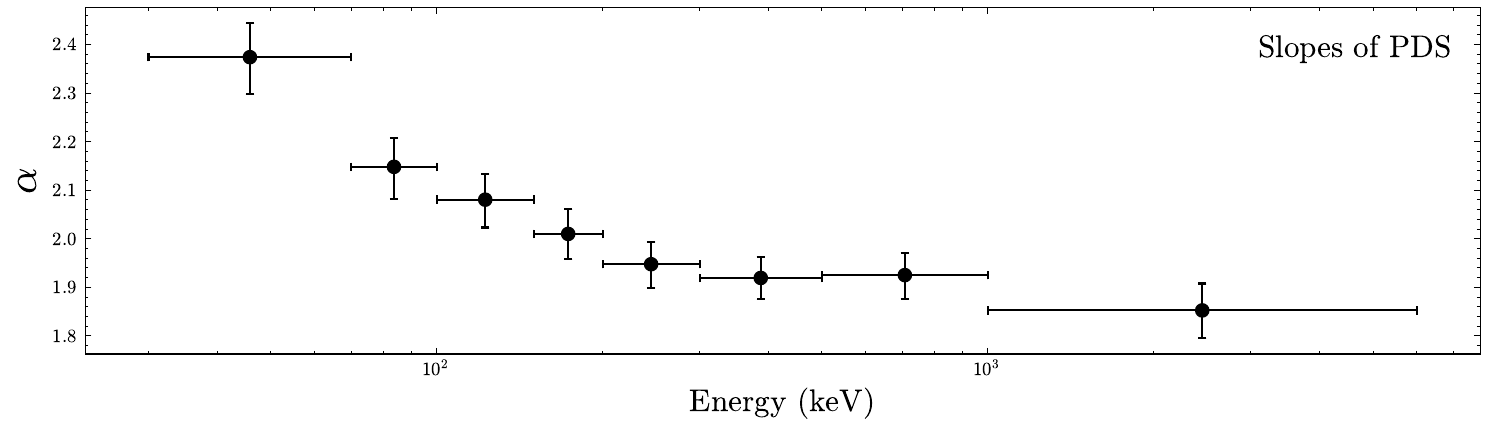}\put(-450, 135){\bf b}
\caption{\textbf{Power density spectra (PDS) of GRB 230307A in various energy bands.} \textbf{Panel (a)} The power density spectra of GRB 230307A in different energy bands are fitted by Eq.(~\ref{pds_bent_pl}). The yellow region represents the 95$\%$ confidence level. \textbf{Panel (b)} The evolution of the slopes of PDS in different energy bands.}
\label{PDS_bands}
\end{figure*}

Another crucial tool used to study the temporal properties of GRB prompt emission is the Fourier power density spectra (PDS), which can shed light on the energy dissipation and geometry of the prompt emission \citep{Beloborodov2000,2016A&A...589A..98G}. As depicted in Figure~\ref{PDS_bands}, 
\begin{equation}
S(f)=A\left[1+\left(\frac{f}{f_\mathrm{b}}\right)^{\alpha}\right]^{-1}+B
\label{pds_bent_pl}
\end{equation}
is used to fit the PDSs in various energy bands for GRB 230307A. The derived slopes $\alpha$ are in the range $\sim 1.85 - 2.37$, which are consistent with the Monte Carlo simulation results generated by the ICMART model \citep{2014ApJ...782...92Z}. There is an evolutionary trend of $\alpha$ decreases as energy increases, particularly in low energy bands \citep{2013MNRAS.431.3608D}. The ``softer wider'' effect can provide an explanation: In lower energy bands, pulses exhibit greater width compared to higher energy bands, resulting in a more pronounced presence of lower frequencies in the former. However, as discussed previously, once the energy band surpasses a certain threshold ($\sim 150$ keV), the predominance of pulse appearance frequency over pulse width results in a diminishing evolution in higher energy bands. The slopes of PDSs are consistent with those derived from Swift GRBs using bent power-law function \citep{2016A&A...589A..98G} (see Extended Data Figure~\ref{PDS_compare}). However, they are a little larger than the values from BATSE GRBs fitted with power-law function \citep{Beloborodov2000}. The main reason is that fitted the slope is higher for bent power-law function than for power-law form. Because in the former model it describes the slope above the break frequency.

\section{DISCUSSION}
\label{sec:discussion}

We found the appearance of 3-dimensional SOC characteristics in the prompt emission phase of GRB 230307A. One-dimensional ($S=1$) SOC characteristics had been discovered in X-ray flares of GRBs, which have been attributed to a magnetic reconnection process \citep{2013NatPh...9..465W,2016ApJS..224...20Y}. The one-dimensional features in the late X-ray flares of GRBs do not conflict with the 3-dimensional features observed during the prompt emission phase in our study. This is attributed to the outward decay of the magnetic field for central engines, resulting in the early activity near the central engine exhibiting three-dimensional characteristics, while the far-away afterglow activity shows one-dimensional features. Polarization studies further support this idea, with high variability during the prompt phase indicating a turbulent 3D magnetic field structure, while stable linear polarization in the afterglow suggests dominant toroidal fields evolving into ordered configurations over time \citep{Lan2019ApJ}. Some works indicate that the central engines of GRB 230307A may be a magnetar (a type of pulsar with a magnetic field strength of $\sim 10^{14}-10^{15}$ Gauss) \citep{2025NSRev..12E.401S}, which can launch Poynting-flux-dominated ejecta to generate luminous prompt emission through magnetic reconnection \citep{2025_mini_jet_Yi,2024MNRAS.529L..67D}. Intense small-scale turbulent magnetic fields can still produce measurable polarization signals as observed in optical data of GRB 160625B \citep{Troja2017}, rather than fully unpolarized emission. From the spectral analysis, the episode showing optical polarization is found to be Poynting-flux-dominated \citep{Zhang2018}, which is consistent with our expectation. For X-ray flares, the magnetic fields in the ejecta will be mainly toroidal, because any radial component decays faster with radius \citep{Giannios2006,2013NatPh...9..465W}. This can naturally explain the different power-law slopes in the duration and peak counts distributions for pulses in the prompt emission and GRB X-ray flares.

Previous efforts have performed statistical analyses of the prompt phase of GRBs \citep{2015ApJ...801...57G,2021FrPhy..1614501L,2023ApJS..265...56L,2024ApJ...965...72M}, leveraging extensive datasets to offer compelling evidence for the existence of SOC characteristics. However, due to the lack of enough pulses in a single GRB, these works compiled pulses in prompt phase of different GRBs, significantly affected by the fluence selection bias and systematic errors. Therefore, it is very important but challenging to determine whether the SOC feature manifests in the prompt phase of individual GRB. The fortuitous confluence of high brightness, broad energy spectrum, and conspicuous multi-peak attributes exhibited by GRB 230307A, coupled with accurate observations from the GECAM satellite \citep{2023GCN.33406....1X}, has afforded us a rich dataset to unveil profound SOC features within this GRB. Our results are closely aligned with theoretical expectations of SOC theory \citep{2014ApJ...782...54A}. This discovery opened a new window to scrutinize the SOC of GRBs and related theoretical inquiries, potentially reshaping our understanding of these enigmatic cosmic phenomena.

\section{CONCLUSIONS}
\label{sec:conclusions}

We have presented a systematic analysis of the temporal statistics of the prompt emission of GRB 230307A using high-quality GECAM observations. The main conclusions are:
\begin{enumerate}
\item The cumulative distributions of pulse peak counts, durations, and waiting times are well described by power-law relations, consistent with the predictions of SOC models.
\item The observed SOC signatures provide evidence supporting magnetic energy dissipation in an ultra-relativistic jet and favor Poynting-flux-dominated GRB emission scenarios.
\item The statistical similarities between GRB 230307A pulses and solar flare events suggest that magnetic reconnection may play an important role in the prompt emission process.
\end{enumerate}
Future observations of additional bright GRBs with high-time-resolution detectors will allow systematic searches for SOC behavior and provide further constraints on the magnetic structure and dissipation physics of relativistic jets.

\begin{acknowledgments}
We are grateful to the anonymous referee for valuable comments and suggestions. We thank the GECAM team for providing the high-energy observations used in this work. This work was supported by the National Natural Science Foundation of China (grant Nos. 12494575, 12273042, 12273009 and 12333007). The GECAM (Huairou-1) mission is supported by the Strategic Priority Research Program on Space Science (Grant No. XDA15360000) of Chinese Academy of Sciences. R. Moradi acknowledges support from the Institute of High Energy Physics of the Chinese Academy of Sciences (E32984U810).
\end{acknowledgments}

\bibliographystyle{aasjournalv7}
\bibliography{sample701}

\appendix
\section{SUPPLEMENTARY MATERIALS: METHODS AND EXTENDED DATA}

\subsection{GECAM Instrument and Data Reduction}
\label{sec:app_methods}

Gravitational wave high-energy Electromagnetic Counterpart All-sky Monitor (GECAM) is a dedicated all-sky gamma-ray monitor constellation funded by the Chinese Academy of Sciences. Currently, it consists of three telescopes: GECAM-A and GECAM-B micro-satellites, which were launched together on December 10, 2020, and GECAM-C (also known as High Energy Burst Searcher, HEBS) aboard the SATech-01 experimental satellite launched on July 27, 2022. Each GECAM telescope is equipped with two types of detectors: Gamma-Ray Detectors (GRDs) and Charged Particle Detectors (CPDs). The GRDs serve as the primary detector of GECAM, each composed of a scintillator and an array of SiPMs. GECAM-A and GECAM-B are equipped with 25 GRDs and 8 CPDs each, while GECAM-C has 12 GRDs and 2 CPDs. All GRDs of GECAM-A and GECAM-B operate in two readout channels: high gain (HG) and low gain (LG), which are independent in terms of data processing, transmission, and dead-time. Most GRDs of GECAM-C follow the same design, with the exception of GRD06 and GRD12, which only operate in high gain mode.

We utilized $GECAMTools$\footnote{https://github.com/zhangpeng-sci/GECAMTools-Public}, the standard software for processing GECAM data, to extract background-subtracted light curves with a time-bin of 5 milliseconds from event-by-event data combined from all 25 GRDs of GECAM-B for higher significance. The background time intervals were set as [$T_0$-50\,s, $T_0$-5\,s] and [$T_0$+160\,s, $T_0$+200\,s]. A zero-order background fitting was applied to interpolate the background during the main prompt phase from $T_0$-0.1\,s to $T_0$+40\,s, which was our main focus. To analyze the temporal behavior in detail, we divided the light curve into eight energy bands: 30-70 keV, 70-100 keV, 100-150 keV, 150-200 keV, 200-300 keV for GECAM-B HG, and 300-500 keV, 500-1000 keV, and 1000-6000 keV for GECAM-B LG. All light curves underwent deadtime correction.

\subsection{Peak Finding and Parameter Extraction}
We utilized the well-established peak search algorithm, $MEPSA$ \citep{2015A&C....10...54G}, specifically designed and calibrated to identify statistically significant local maxima in GRB light curves, to pinpoint peaks in the background-subtracted light curve. Each pulse candidate is characterized by its peak time, peak counts rate, and the timescale at which the pulse was identified with the highest signal-to-noise ratio ($S/N$). The results show that the $S/N$ of all pulse candidates is larger than 5. The three pieces of information of pulses identified by $MEPSA$ were selected for our research:
\begin{enumerate}
    \item PeakT: peak time, defined as $t_{\text{peak} }$ in this research;
    \item BinT: detection timescale;
    \item PeakR: peak counts rate, representing the counts rate of the bin at the peak time.
\end{enumerate}
The Parameters in our research are processed as follows:
\begin{enumerate}
    \item Waiting time ($\Delta t$): defined as $\Delta t=t_{\text{peak},i+1}-t_{\text{peak},i}$, where $t_{\text{peak},i+1}$ and $t_{\text{peak},i}$ are the peak times for the $(i + 1)$th and $i$th pulse, as the time-scale for waiting time;
    \item Duration ($T$): define BinT as the time-scale for duration;
    \item Peak counts ($P$): defined as the peak counts rate multiplied by the time resolution (5 ms), as the flux-scale for peak flux.
\end{enumerate}

\subsection{Cumulative Distribution Fitting}
Due to the limited number of samples, cumulative distributions are utilized for the three parameters, such as peak counts $P$, duration $T$ and waiting time $\Delta t$, which can offer a more effective approach than differential distributions. The cumulative distribution ($N_{\mathrm{cum}}(>x)$) represents the number of events larger than a specific value $x$. The uncertainty of the cumulative distribution is estimated as  $\sigma_{\mathrm{cum,i}}=\sqrt{N_{\mathrm{cum,i}}-N_{\mathrm{cum,i+1}}}$ \citep{2019ApJ...880..105A}, which takes only independent events into account for the estimation of the uncertainty.

Identifying the pulses and determining the waiting time between adjacent pulses of a gamma-ray burst is generally challenging due to the uncertainty surrounding the presence of smaller pulses within this interval. Fortunately, we have sufficient number of data available for statistical analysis. Initially, we utilized the bent power-law function \citep{2016A&A...589A..98G}
\begin{equation}
N_{\mathrm{cum}} (>x) = A\left[1+\left(\frac{x}{x_\mathrm{b}}\right)^{\beta}\right]^{-1}
\label{Tpw_bent_pl}
\end{equation}
to fit the distributions and determine a break point, denoted as $x_{\rm b}$, which serves as a measure of the resolution in identifying pulses. This definition differs from the estimate in the previous study \citep{2015ApJ...814...19A} (where $x_0$ is set as the peak in the event count histogram). The data above $x_{\rm b}$ are considered a complete sample. The same method is applied to the duration ($T$) and peak counts. Subsequently, we fitted the cumulative distributions above $x_{\rm b}$ using Eq.~\ref{cum_pl}, which is derived from the integral of a simple power-law function (Eq.~\ref{diff_pl}) commonly used for a differential distribution. The parameters of interest in our fitting are $\alpha_x$ from Eq.~(\ref{cum_pl}), as well as $x_{\rm b}$ from Eq.~(\ref{Tpw_bent_pl}).

We assess the goodness of fit using the uncertainties $\sigma_{\text{cum,i}}$ associated with the values $N_{\text{cum,i}}$ in the fitted distribution through the standard reduced chi-square ($\chi^2$) criterion \citep{2015ApJ...814...19A,Cheng2020}
\begin{equation}
\chi_{\text{cum}}=\sqrt{\frac{1}{(n_x-n_{\text{par}})}\sum_{i=1}^{n_x}\frac{\left[N_{\text{cum,th}}(x_i)-N_{\text{cum,obs}}(x_i)\right]^2}{\sigma_{\text{cum},i}^2}},
\end{equation}
for the cumulative distribution function, where $n_{\text{x}}$ is the number of data points used in the fitting, $n_{\text{par}}$ is the number of free parameters, $N_{\text{cum,obs}}(x_{i})$ denotes the observed values, and $N_{\text{cum,th}}(x_{i})$ represents the theoretical values for the cumulative distribution. The fitting process is implemented using the Python module \emph{emcee} for cumulative distribution. The fitting results are listed in Table~\ref{tab:fitting_results}. In Extended Data Figure~\ref{Residual}, the fit residuals are shown. The residuals are consistent with zero in $1\sigma$ confidence level.

\subsection{Simulations of Poisson and Cellular Automaton Processes}
Here, the non-stationary Poisson process is used to explain the waiting time distribution. To test the robustness of the results, a Poisson process simulation was performed to test if a Poisson process can account for the observed waiting time distribution. In this simulation, we first generated a set of burst arrival times within a 40-s interval following a Poisson process. To further test the robustness of the result, we also explored several different burst rates. Next, to further account for the possible effects of pulse pile-up, we assigned a finite width to each pulse and allowed the pulses to overlap with one another. Finally to consider the possible temporal modulation of the central machine adopted a peak-flux model that first rises as exponential and then decay as a power-law function with power index $-2.7$ \citep{Chincarini2010}. An example of the simulated light curves are shown in Extended Data Figure~\ref{example_simulation}. After the simulation, we then applied the same MESPA peak-finding algorithm used for the observational data to identify bursts in this simulated sequence, and we used the detected bursts to compute the waiting time distribution. We also directly used the burst arrival-time sequence generated by the Poisson process to construct the waiting-time distribution for further comparison. As shown in Extended Data Figure~\ref{Poisson_simulation}, our results implies that if GRB pulses were driven by a Poisson process, the resulting waiting time distribution would be exponential, which fails to reproduce the observed waiting time distribution. This discrepancy is particularly pronounced at longer waiting times. Moreover, observational effects—such as the temporal decline of burst energy and the MESPA peak-finding procedure—further exacerbate this discrepancy.

Additionally, the cellular automaton simulation was also employed to further compare the result given by Poisson process and SOC, which is widely used in the SOC framework \citep{1987PhRvL..59..381B,1991ApJ...380L..89L,Norman2001,Wang2023}. The simulation procedure is shown as follows. We consider a three-dimensional regular cubic lattice, with each lattice node represented by three integers, ${i,j,k}$. The magnetic field of each lattice node is given by $B_{\rm i, j, k}$. The magnetic field is assumed to be zero at the boundaries \citep{Norman2001}. Small perturbations $\delta B$ are added to randomly selected interior nodes. After each addition, the $B$ difference between each cubic lattice and six neighbors can be calculated as
\begin{equation}
\Delta B_{\rm i, j, k} = B_{\rm i, j, k} - \frac{1}{6}\sum_{\rm nn}B_{\rm nn},
\end{equation}
where $B_{\rm nn}$ is the difference between two nearest neighbors. At each time step, random perturbations $\delta B$ are randomly added to the cubic lattice. Then we compare $\Delta B_{\rm i, j, k}$ with a specified threshold $B_c$. In the SOC model, the perturbation $\delta B$ needs to be smaller than the specified threshold $B_c$.
If $\Delta B_ {\rm i,j,k}$ is less than $B_c$, the system will keep stability. The avalanches occur when $\Delta B_ {\rm i,j,k}$ is larger the specified threshold $B_c$. Then the $B$ field is redistributed according to the redistribution rule:
\begin{equation}
B_{\rm i, j, k}' = B_{\rm i, j, k} - \left(\frac{6}{7} \right)s B_c,
\end{equation}
where $s = \Delta B_{\rm i, j, k} / \mid \Delta B_{\rm i, j, k} \mid$. After the redistribution of the $B$ field, random perturbations are continuously added to the lattice at each time step. The next avalanche will occur when the threshold is reached.

Firstly, we assume that the central engine's unsteady behavior is a Poisson process, i.e., a succession of shells randomly ejected by the central engine \citep{Kobayashi1997,Daigne1998,Nakar2002}. In our model, the randomly shells correspond to that the perturbation $\delta B$ is added randomly \citep{1991ApJ...380L..89L}. The perturbation $\delta B$ is chosen from a random distribution in the range [-0.2, 0.8] \citep{Wang2023}. Simulations are carried out on a 3D lattice with $10^6$ iterations. The threshold value is set as $B_c=7$. We recorded the time between adjacent avalanches and analyze the distribution of waiting times after accumulating enough simulated data. As shown in Extended Data Figure~\ref{CA}, the distribution of waiting time for random driving force follows an exponential distribution and the distribution of waiting time for driving force related to the random walk follows a power-law distribution.

\subsection{Power Density Spectra}
Power density spectra (PDS) of GRBs provide useful information on GRBs, indicating their self-similar temporal structure \citep{Beloborodov2000}. We used the $Stingray$ package to compute PDS from the background-subtracted light curve. The PDS is normalized using Leahy normalization. In order to provide a detailed description of the PDS, a modified Bent Power-law function Eq.~\ref{pds_bent_pl} is used to model the plateaus at low and high frequencies, as well as the slopes \citep{2016A&A...589A..98G}. It is important to note that the segment with frequencies lower than 0.1\,Hz was excluded during the fitting of the PDS. This fitting process is implemented using the Python module \emph{emcee} with $\mathrm{ln}\mathcal{L}=\sum_i\ln\left(\frac{\exp({-y_i/y_{\text{th}}})}{y_{\text{th}}}\right)$ for PDS \citep{2010MNRAS.402..307V}, where $y_i$ denotes the observed values, and $y_{\text{th}}$ represents the theoretical values for Eq.(~\ref{pds_bent_pl}).

\subsection{Detailed Explanation of $\alpha_{T}$, $\alpha_{P}$ and $\alpha_{\Delta t}$}
\begin{figure*}[htp]
\centering
\includegraphics[width=0.7\textwidth, angle=0]{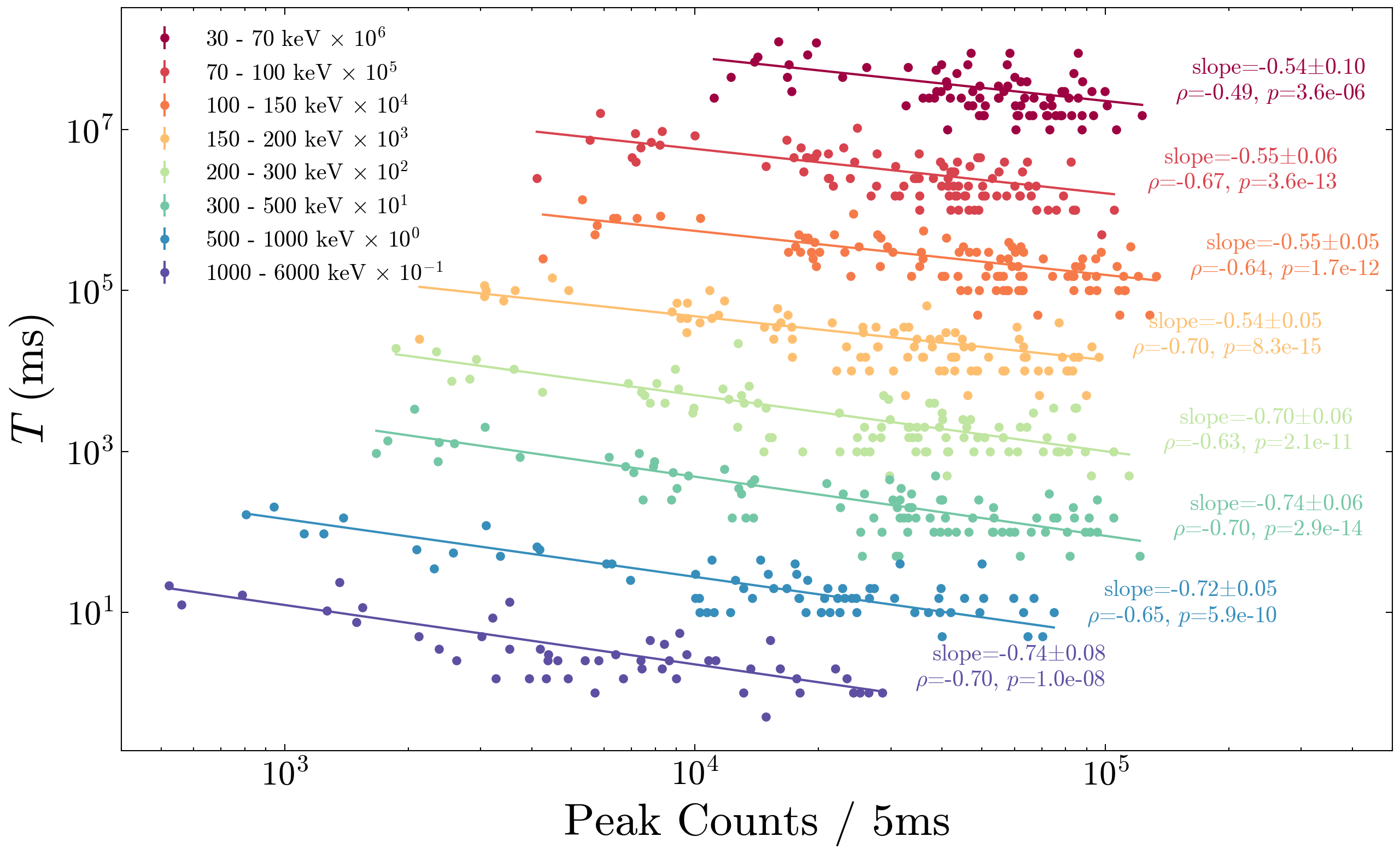}\put(-315, 205){\bf a}\\
\includegraphics[width=0.7\textwidth, angle=0]{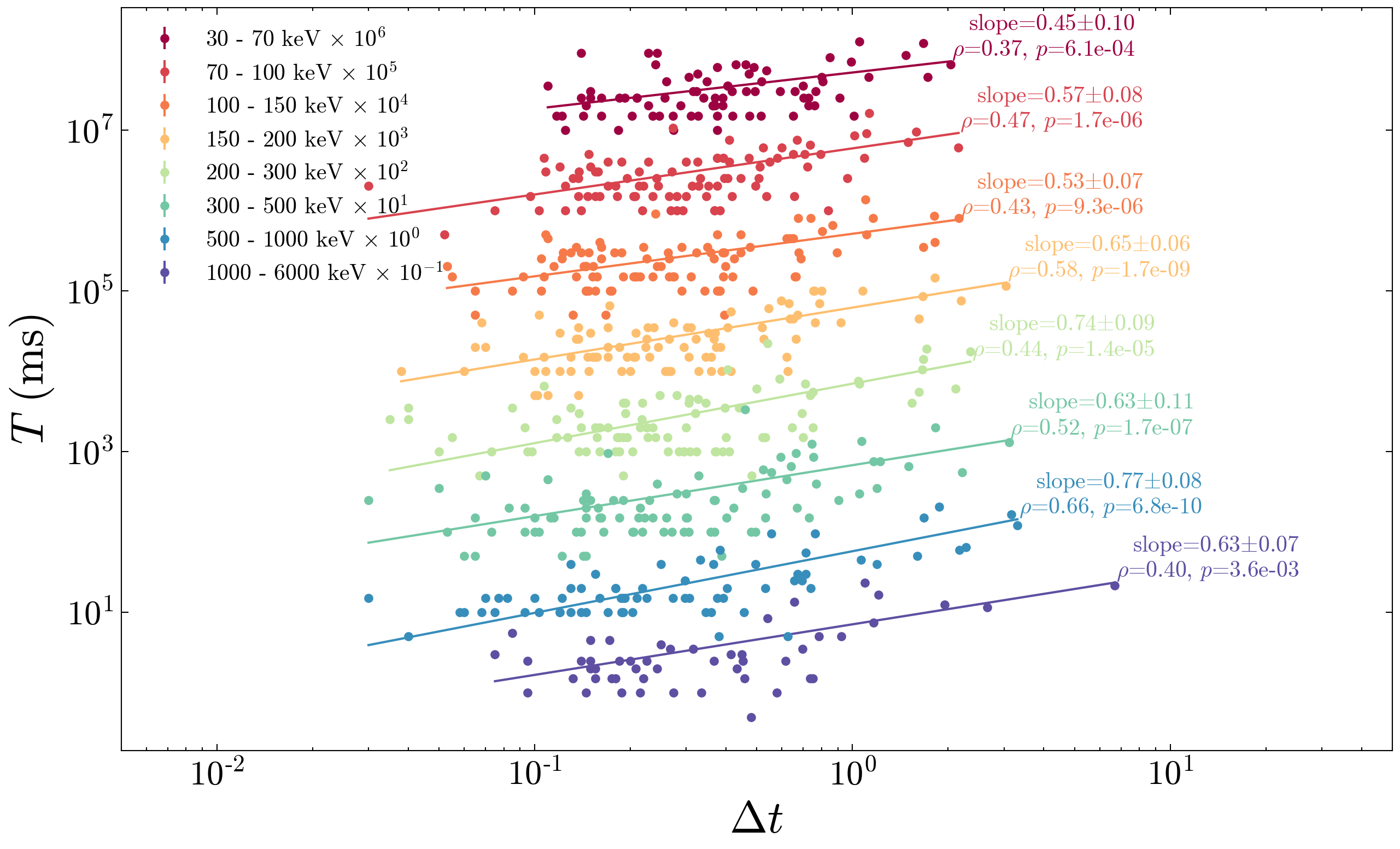}\put(-315, 205){\bf b}
\caption{\textbf{Correlations of waiting time ($\Delta t$), duration ($T$), and peak counts ($P$) across various energy bands.}
\textbf{Panel (a)} Correlation between duration ($T$) and peak counts ($P$).
\textbf{Panel (b)} Correlation between duration ($T$) and waiting time ($\Delta t$).
Solid lines (colored accordingly) represent the trend fits obtained using a simple power-law function. Spearman correlation coefficient and $p$-value are calculated to measure the relationship between two parameters, highlighted in corresponding colors in two panels. The Spearman coefficient measures the strength and direction of monotonic association (ranging from -1 to 1, where values closer to -1 represent strong anti-correlation and values closer to 1 represent strong positive correlation), while the $p$-value indicates the significance of the correlation, with smaller $p$-values representing stronger significance.}
\label{Corr}
\end{figure*}

As shown in \textbf{panel a} of Figure~\ref{Corr}, we observe an anti-correlation between the duration ($T$) and peak counts ($P$), suggesting that brighter bursts are typically narrower and the identified pulses are dominated by brighter and narrower bursts, across all energy bands. Notably, in high energy bands, the slope of the anti-correlation steepens. These wider bursts typically have longer waiting time as shown in \textbf{panel b} of Figure~\ref{Corr}. We now apply these two correlations to compare the observed values of $\alpha_{T}$, $\alpha_{P}$ and $\alpha_{\Delta t}$ with their theoretical predictions.

\begin{figure*}[htp]
\centering
\includegraphics[width=0.34\textwidth]{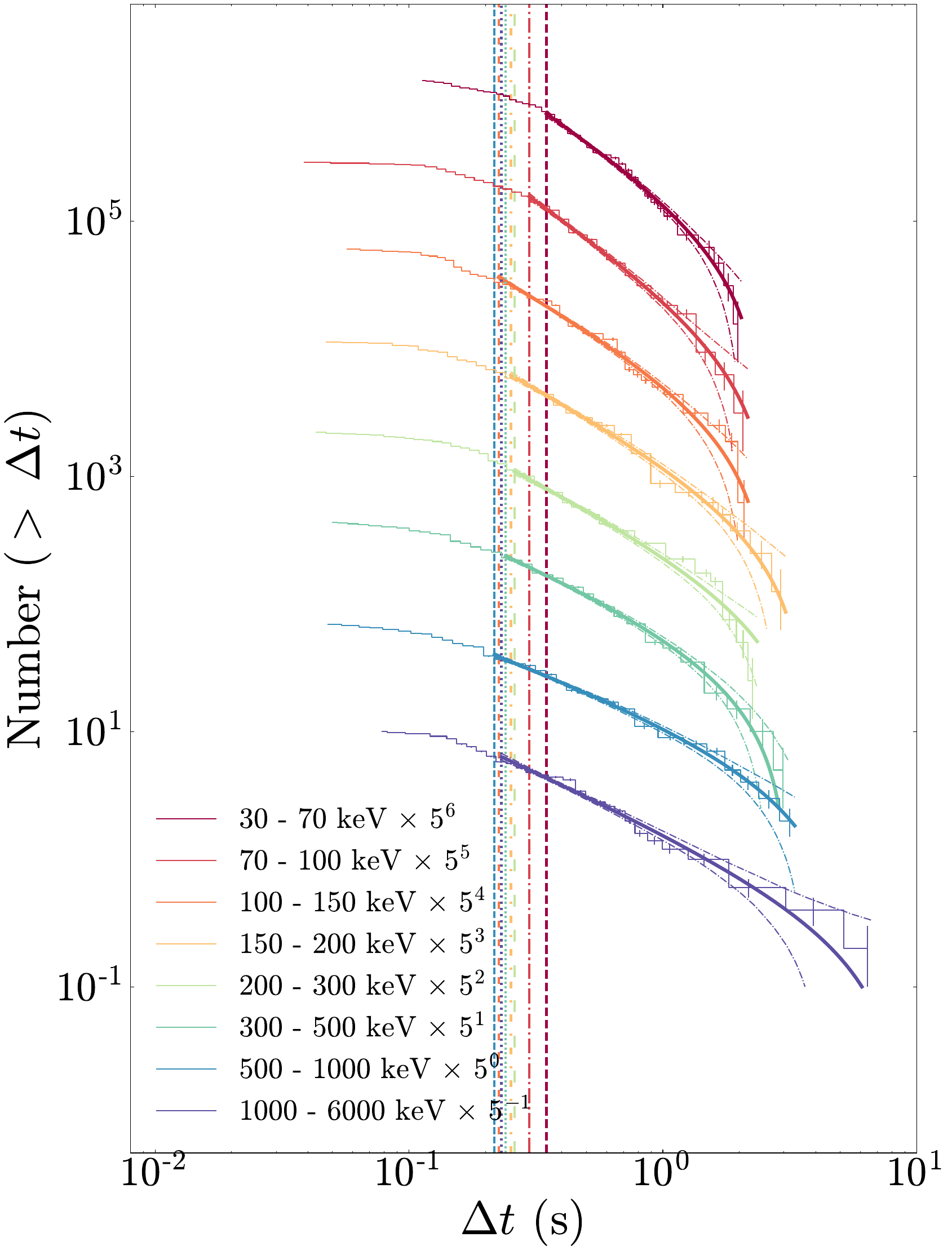}\put(-140, 210){\bf a}
\includegraphics[width=0.33\textwidth, angle=0]{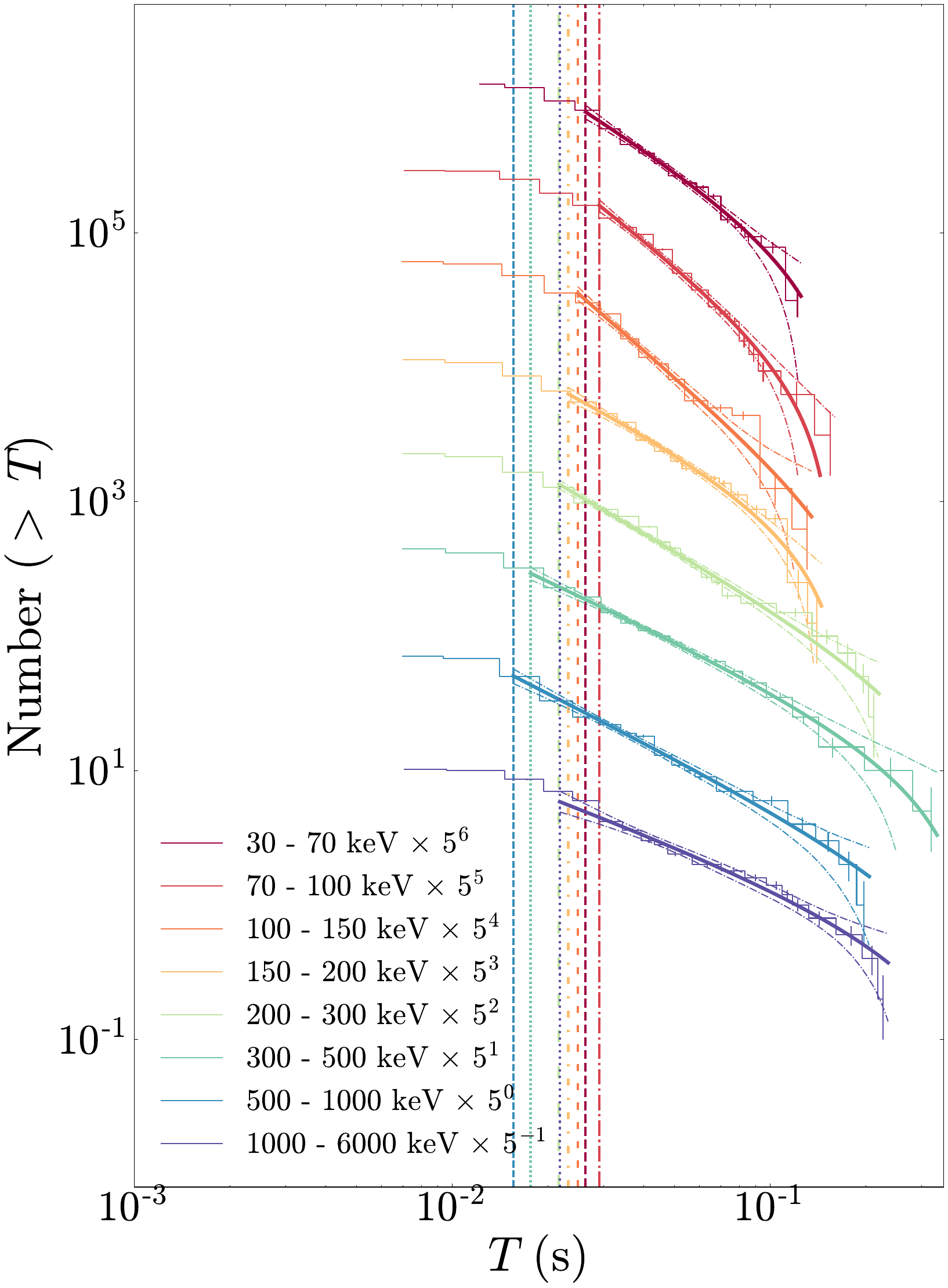}\put(-135, 210){\bf b}
\includegraphics[width=0.33\textwidth, angle=0]{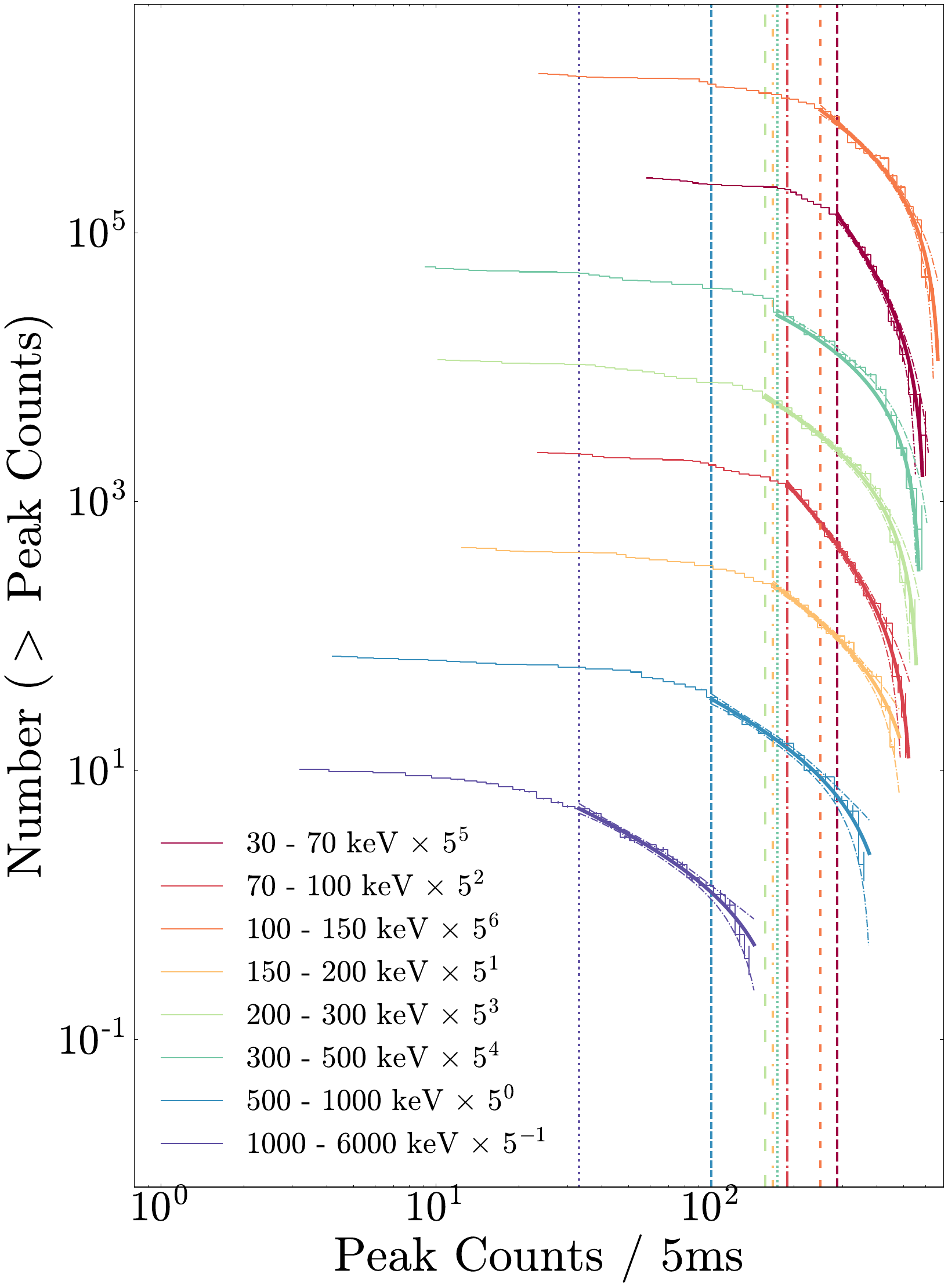}\put(-135, 210){\bf c}\\
\includegraphics[width=0.33\textwidth, angle=0]{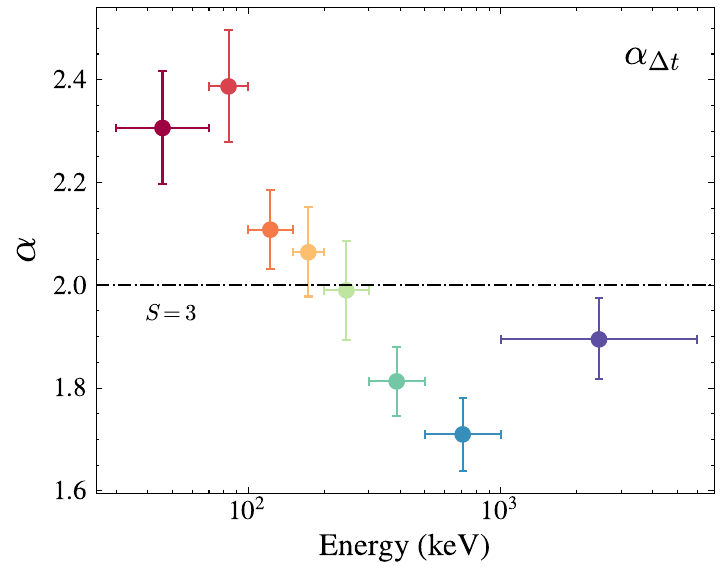}\put(-155, 135){\bf d}
\includegraphics[width=0.33\textwidth, angle=0]{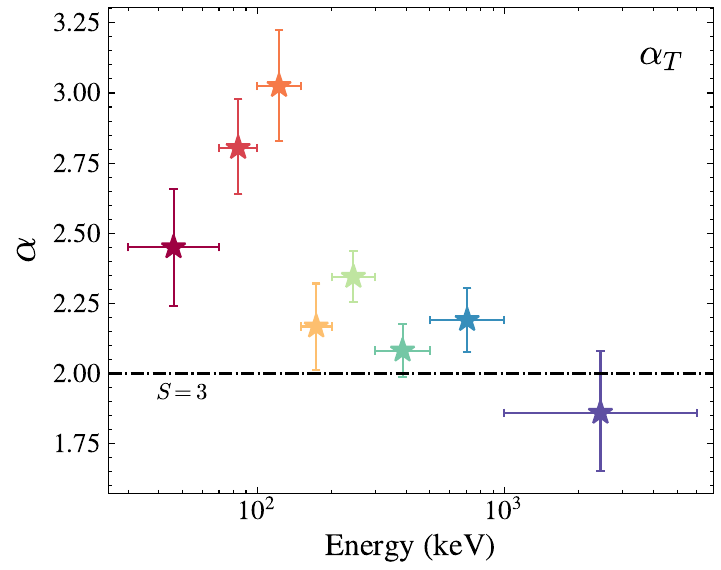}\put(-155, 135){\bf e}
\includegraphics[width=0.33\textwidth, angle=0]{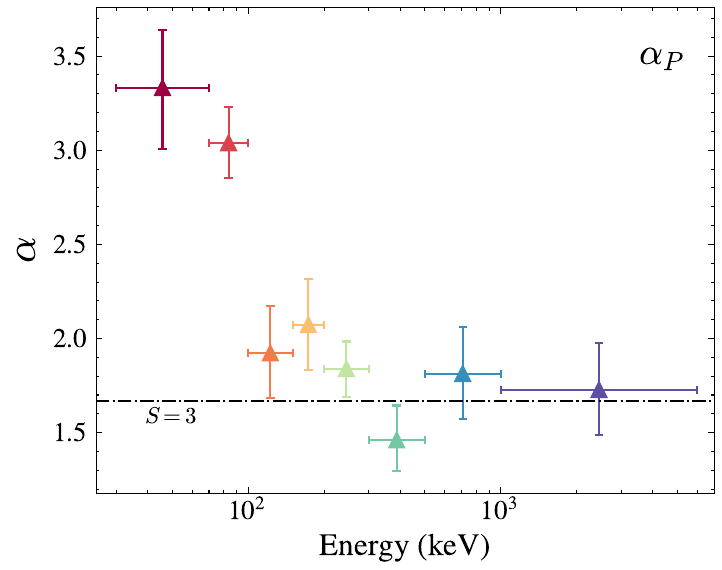}\put(-155, 135){\bf f}\\
\caption{\textbf{Waiting time ($\Delta t$), duration ($T$), and peak counts ($P$) distributions of pulses in various energy bands.} 
\textbf{Panel (a)} The cumulative distributions of $\Delta t$ of pulses in various energy bands are shown in stepwise lines. The vertical dotted lines indicate the break $x_{\rm b}$ in the distributions, above which they are fitted. The best fit using Eq.~(\ref{cum_pl}) is presented as solid line with the 95$\%$ confidence interval between dashed lines.
\textbf{Panel (b)} Same as panel (a), but for the $T$.
\textbf{Panel (c)} Same as panel (a), but for the $P$.
\textbf{Panel (d)} The evolution of the power-law slopes for $\Delta t$ (dots) as a function of energy range. The dashed line shows the prediction (2.0) of the 3-dimension SOC model. 
\textbf{Panel (e)} and \textbf{Panel (f)} are same as panel (d), but for the $T$ and $P$, respectively. As the energy increases, the slope for $\Delta t$ (panel d) exhibits a temporary pause at $\sim$150 keV but continues to decrease at higher energies up to $\sim$1000 keV and beyond; in contrast, the slopes for $T$ (panel e) and $P$ (panel f) level off and remain nearly constant from $\sim$150 keV upward.}
\label{Tpw_bands}
\end{figure*}

\begin{figure*}[htp]
\centering
\includegraphics[width=0.7\textwidth, angle=0]{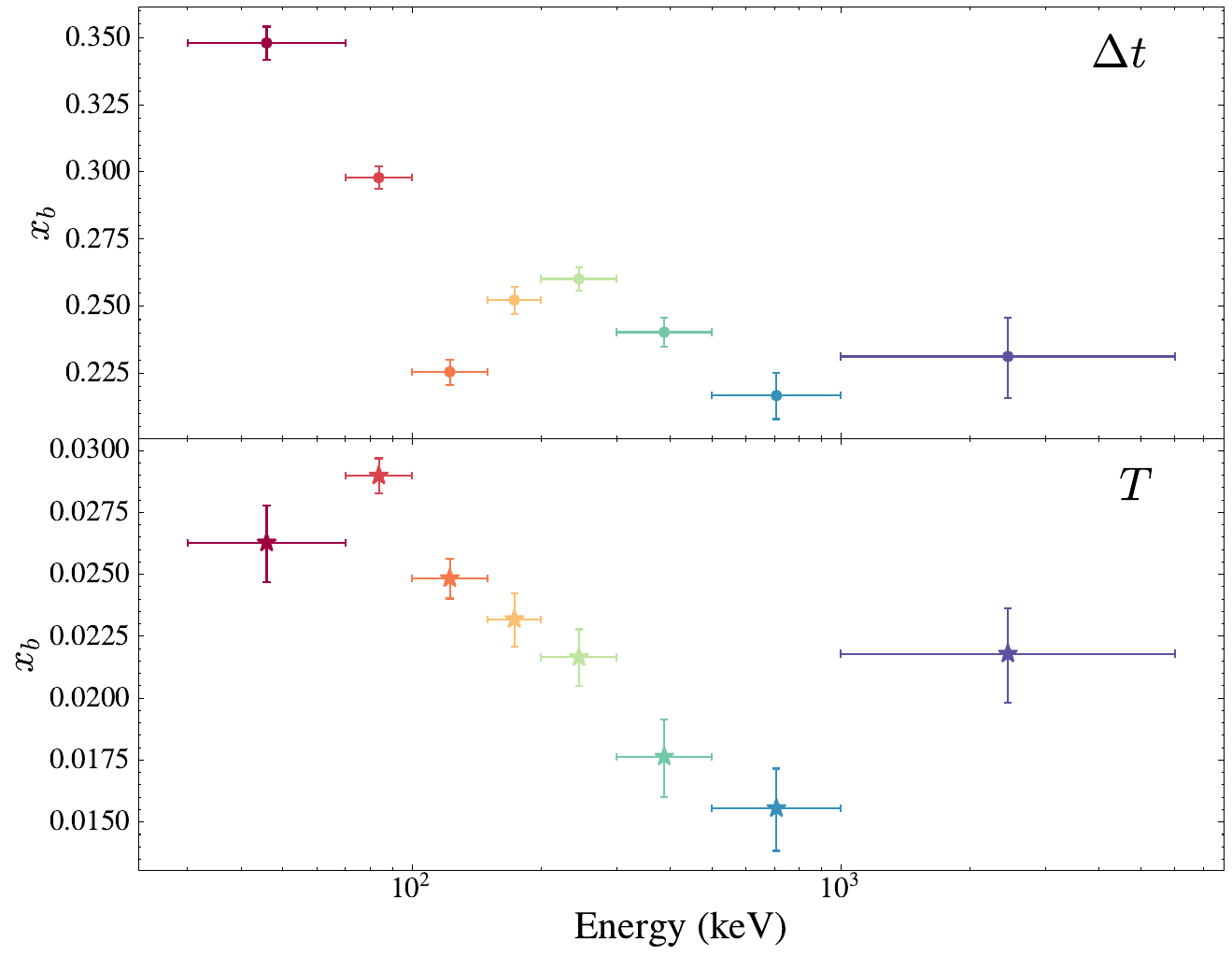}\\
\caption{\textbf{The evolution of break point ($x_{\rm b}$, in seconds) of $\Delta t$ and $T$ as a function of energy.} The upper panel depicts the stabilization of the break points of $\Delta t$ (dots) in higher energy bands. The lower panel illustrates a decreasing trend in the break points of $T$ (star) as the energy band increases from low to high.}
\label{Tpw_bent}
\end{figure*}

\paragraph{For Peak Counts ($P$)} From Figure~\ref{Corr} {\bf a/b}, we can infer that the brightest bursts are typically the narrowest and also have the shortest waiting time. However, as shown in Table~\ref{tab:fitting_results} and Figure \ref{Tpw_bent} {\bf a}, the break times in the two lowest energy bands are significantly longer than those in the higher energy bands. It implies that the brightest bursts in the two energy bands, i.e. with the shortest duration and waiting time, may be missed more likely in identification process. Therefore, it will naturally result in a deficit in identifying the brightest bursts in the two lowest energy bands, resulting in much steeper power-law slope $\alpha_{P}$ than the theoretically predicted value of 1.67, as shown in Figure~\ref{Tpw_bands} {\bf e}. In the higher energy bands, the above selection effect does not exist, and consequently the observed $\alpha_{P}$ is consistent with 1.67 predicted by 3-dimensional SOC theory.

\paragraph{For Duration ($T$)} From \textbf{panel d} of Figure~\ref{Tpw_bands}, two prominent characteristics regarding $\alpha_{T}$ are observed. First, in the low-energy region, $\alpha_{T}$ significantly deviates from 2 (greater than 2). Second, in the high-energy region, $\alpha_{T}$ fluctuates around and are consistent with 2. This deviation in the low-energy bands can be attributed to observational effects: the ``softer wider'' effect tends to make the low-energy pulses to be wider. Observationally it is more difficult to identify wide and weak pulses, resulting in detection deficit of low-energy wide pulses, which leads to a steeper power-law slope of the cumulative distribution of duration, i.e. $\alpha_{T}$ becomes greater than 2. Conversely, this effect is absent in the high-energy region, where $\alpha_{T}$ is consistent with 2, also in accordance with our observations.

\paragraph{For Waiting Time ($\Delta t$)} A slow decreasing trend in $\Delta t$ around 2 can be found from \textbf{panel d} of Figure~\ref{Tpw_bands}: $\alpha_{\Delta t}>2$ below about 200 keV and $\alpha_{\Delta t}<2$ above about 200 keV. According to Table~\ref{tab:fitting_results}, the number of pulses in the low-energy bands is larger than that in the high-energy bands. A pulse can occur in all energy regions, and its non-detection in a particular energy region is solely due to selection effects. There are two selection effects: (i) The scarcity of pulses in the high-energy region is evidently due to missing some pulses identified in the low-energy region. Missing these pulses inevitably leads to an increase in the measured pulse intervals, resulting in $\alpha_{\Delta t}<2$; (ii) From Figure~\ref{Corr} \textbf{b}, wider bursts typically have longer waiting time. As discussed above for duration, the detection deficit should appear for wider pulses in lower energy bands, resulting in reduction of identified bursts with a longer waiting time, i.e., $\alpha_{\Delta t}>2$.

\clearpage
\begin{table*}[htp]
\centering
\scriptsize
\caption{\textbf{Fitting results of waiting time ($\Delta t$), duration ($T$) and peak counts ($P$) distributions.} All errors represent the 1$\sigma$ uncertainties.}
\label{tab:fitting_results}
\begin{threeparttable}
\begin{tabular}{c|c|ccc|ccc|ccc}
\hline
\hline
Energy Range& \multicolumn{1}{|c|}{Pulse Number}& \multicolumn{3}{|c|}{Waiting Time ($\Delta t$)} & \multicolumn{3}{|c|}{Duration ($T$)} & \multicolumn{3}{|c}{Peak Counts ($P$)}\\
\cline{3-11}
(keV) & $N$ & $x_{\rm b}$ (ms)\tnote{$\ast$} & $\alpha$\tnote{$\star$} & $\chi^2$\tnote{$\dagger$} &  $x_{\rm b}$ (ms)\tnote{$\ast$} & $\alpha$\tnote{$\star$} & $\chi^2$\tnote{$\dagger$} & $x_{\rm b}$\tnote{$\ast$} & $\alpha$\tnote{$\star$} & $\chi^2$\tnote{$\dagger$} \\
\hline
30-70 & 82 & $348^{+6}_{-6}$ & $2.31^{+0.11}_{-0.11}$ & 0.80 & $26^{+2}_{-2}$ & $2.45^{+0.21}_{-0.21}$  & 0.40 & $286^{+2}_{-2}$ & $3.33^{+0.31}_{-0.32}$ & 1.05 \\
70-100 & 93 &  $298^{+4}_{-4}$ & $2.39^{+0.11}_{-0.11}$ & 0.53 & $29^{+1}_{-1}$ & $2.80^{+0.17}_{-0.17}$ & 0.67 & $189^{+2}_{-2}$ & $3.04^{+0.19}_{-0.18}$ & 0.54 \\
100-150 & 98 & $226^{+5}_{-5}$ & $2.11^{+0.08}_{-0.08}$ & 1.48 & $25^{+1}_{-1}$ & $3.03^{+0.20}_{-0.20}$ & 0.98 & $249^{+2}_{-2}$ & $1.92^{+0.25}_{-0.24}$ & 2.13 \\
150-200 & 91 &  $252^{+5}_{-5}$ & $2.06^{+0.09}_{-0.09}$ & 0.79 & $23^{+1}_{-1}$ & $2.17^{+0.15}_{-0.15}$ & 0.65 & $167^{+2}_{-2}$ & $2.07^{+0.24}_{-0.24}$ & 0.66 \\
200-300 & 91 & $260^{+4}_{-4}$ & $1.99^{+0.10}_{-0.10}$ & 0.78 & $22^{+1}_{-1}$ & $2.34^{+0.09}_{-0.09}$ & 1.13 & $157^{+2}_{-2}$ & $1.84^{+0.15}_{-0.15}$ & 0.53 \\
300-500 & 89 & $240^{+5}_{-5}$ & $1.81^{+0.07}_{-0.07}$ & 0.56 & $18^{+2}_{-2}$ & $2.08^{+0.09}_{-0.09}$ & 0.21 & $174^{+2}_{-2}$ & $1.46^{+0.18}_{-0.16}$ & 1.25 \\
500-1000 & 72 & $217^{+9}_{-9}$ & $1.71^{+0.07}_{-0.07}$ & 0.64 & $16^{+2}_{-2}$ & $2.19^{+0.11}_{-0.11}$ & 0.41 & $100^{+2}_{-2}$ & $1.81^{+0.25}_{-0.24}$ & 0.34\\
1000-6000 & 51 & $231^{+14}_{-15}$ & $1.89^{+0.08}_{-0.08}$ & 0.99 & $22^{+2}_{-2}$ & $1.86^{+0.22}_{-0.21}$ & 0.66 & $33^{+1}_{-1}$ & $1.73^{+0.25}_{-0.24}$ & 0.49\\
\hline
\end{tabular}
    \begin{tablenotes}
        \footnotesize
        \item[$\ast$] Break points of cumulative distributions.
        \item[$\star$] Slopes of the power-law distributions.
        \item[$\dagger$] Goodness-of-fit results.
    \end{tablenotes}
\end{threeparttable}
\end{table*}

\clearpage
\setcounter{figure}{0}
\renewcommand{\figurename}{Extended Data Fig.}

\begin{figure*}[htp]
\centering
\includegraphics[width=\textwidth, angle=0]{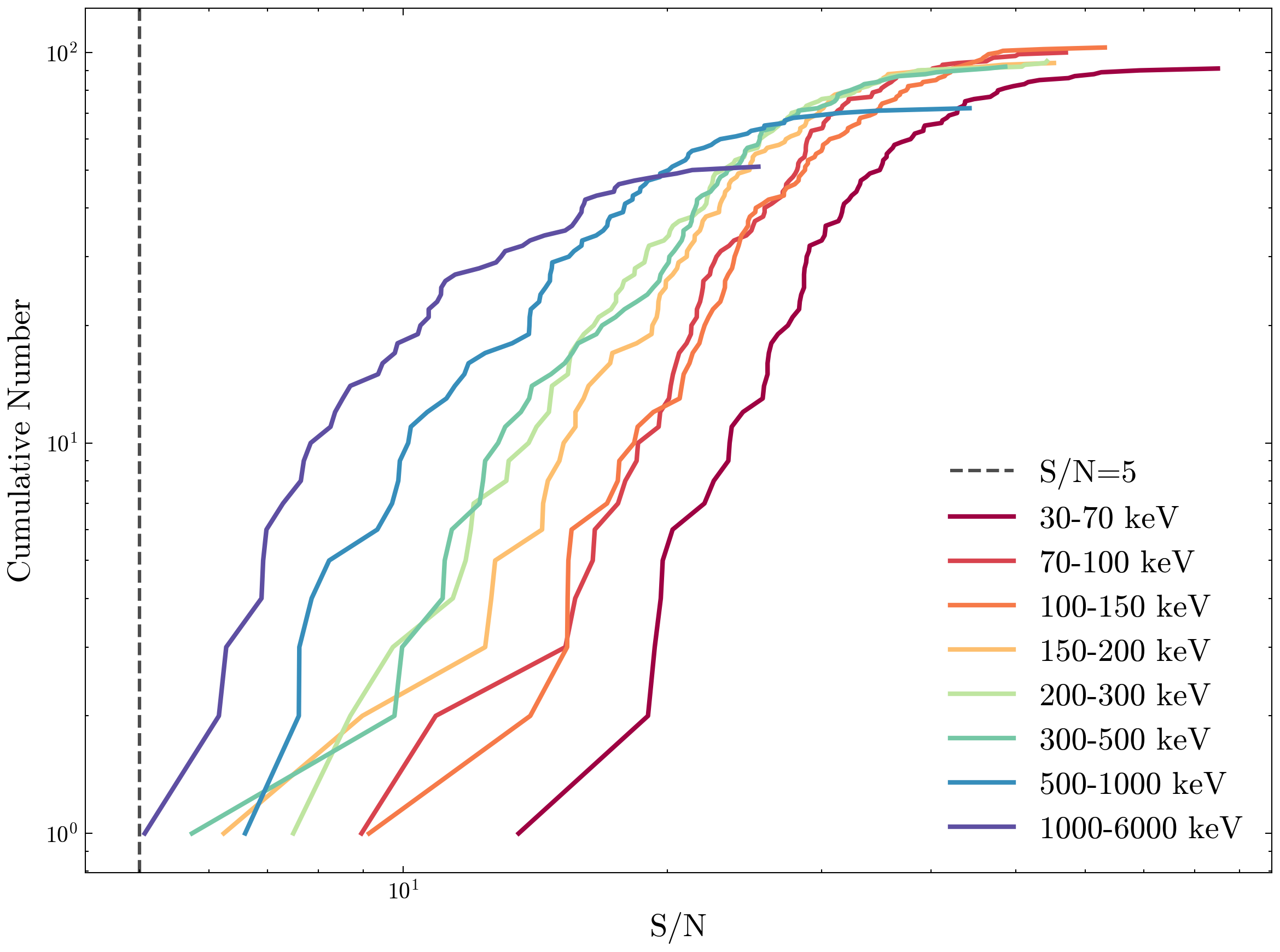}
\caption{\textbf{S/N of pulses in each band of GRB 230307A identified by MEPSA.}}
\label{SNR}
\end{figure*}

\begin{figure*}[htp]
\centering
\includegraphics[width=0.33\textwidth, angle=0]{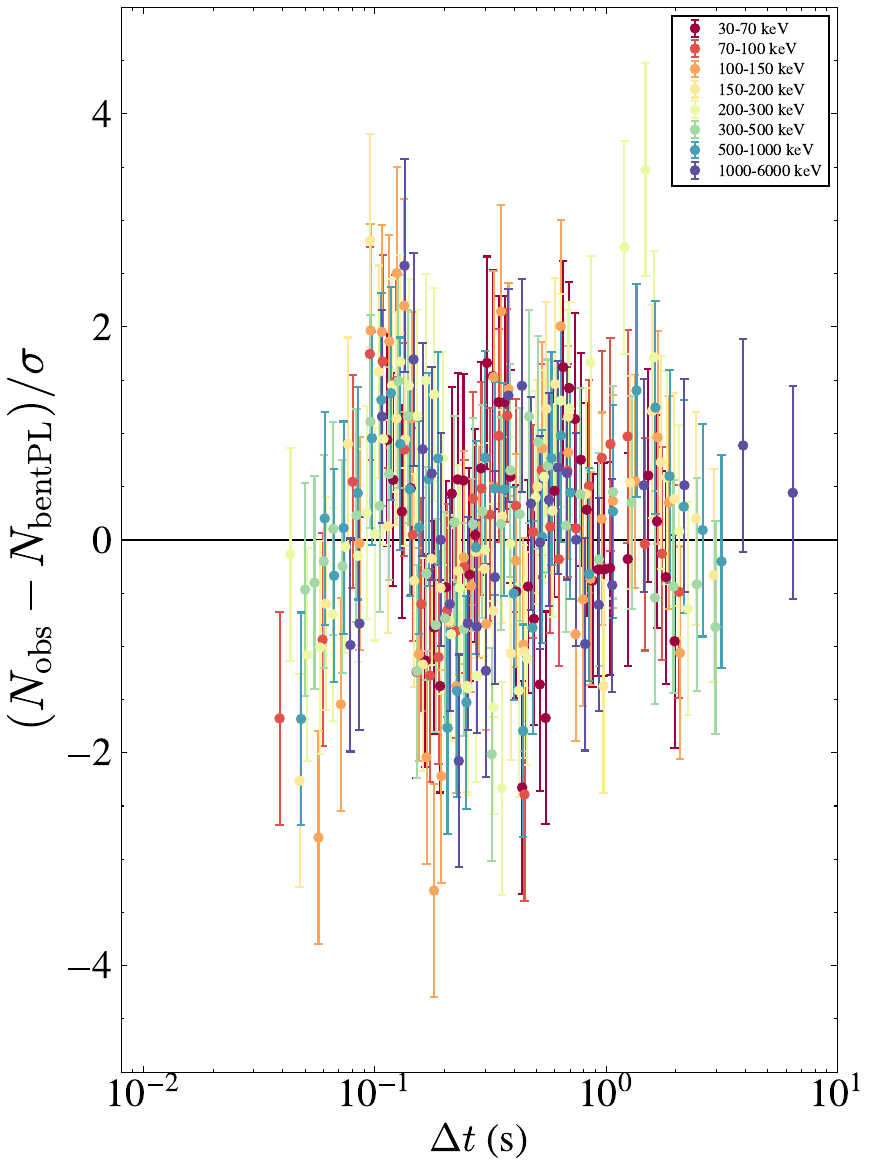}\put(-160, 225){\bf a}
\includegraphics[width=0.33\textwidth, angle=0]{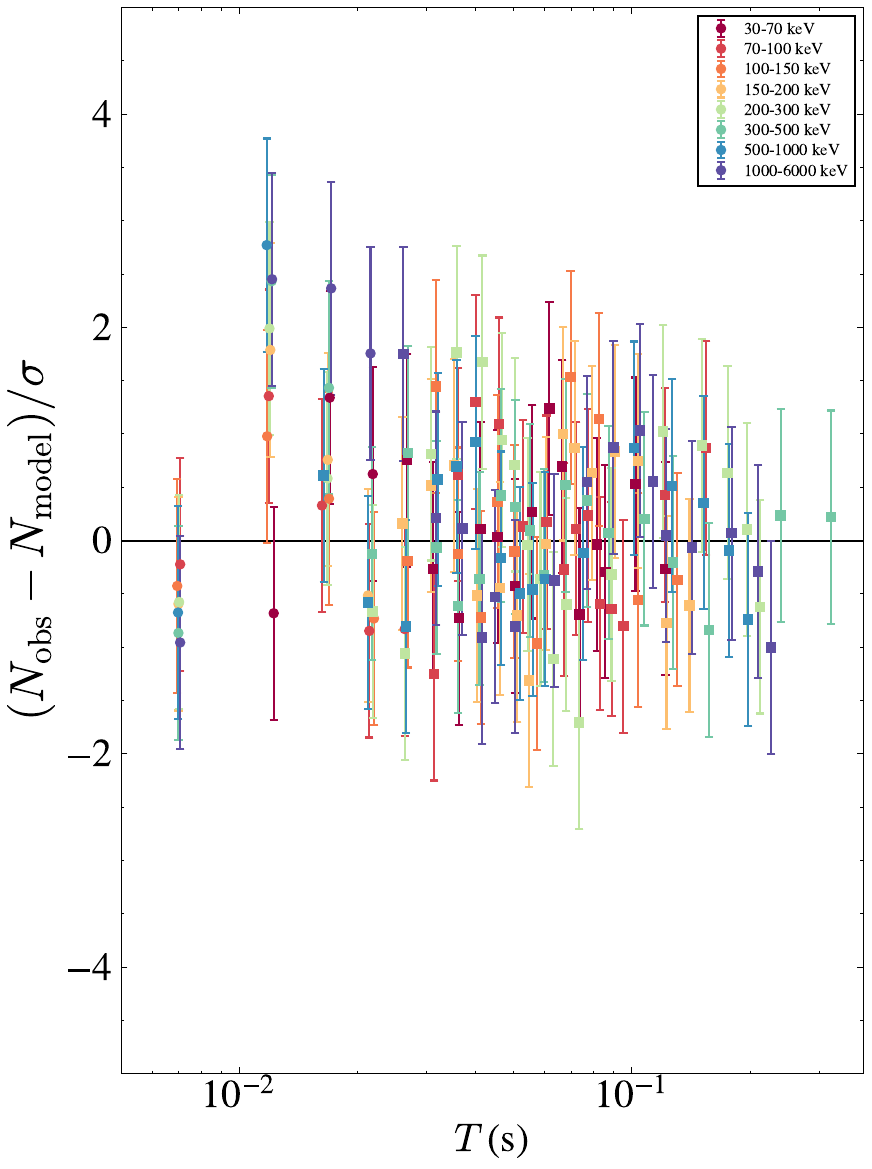}\put(-160, 225){\bf b}
\includegraphics[width=0.33\textwidth, angle=0]{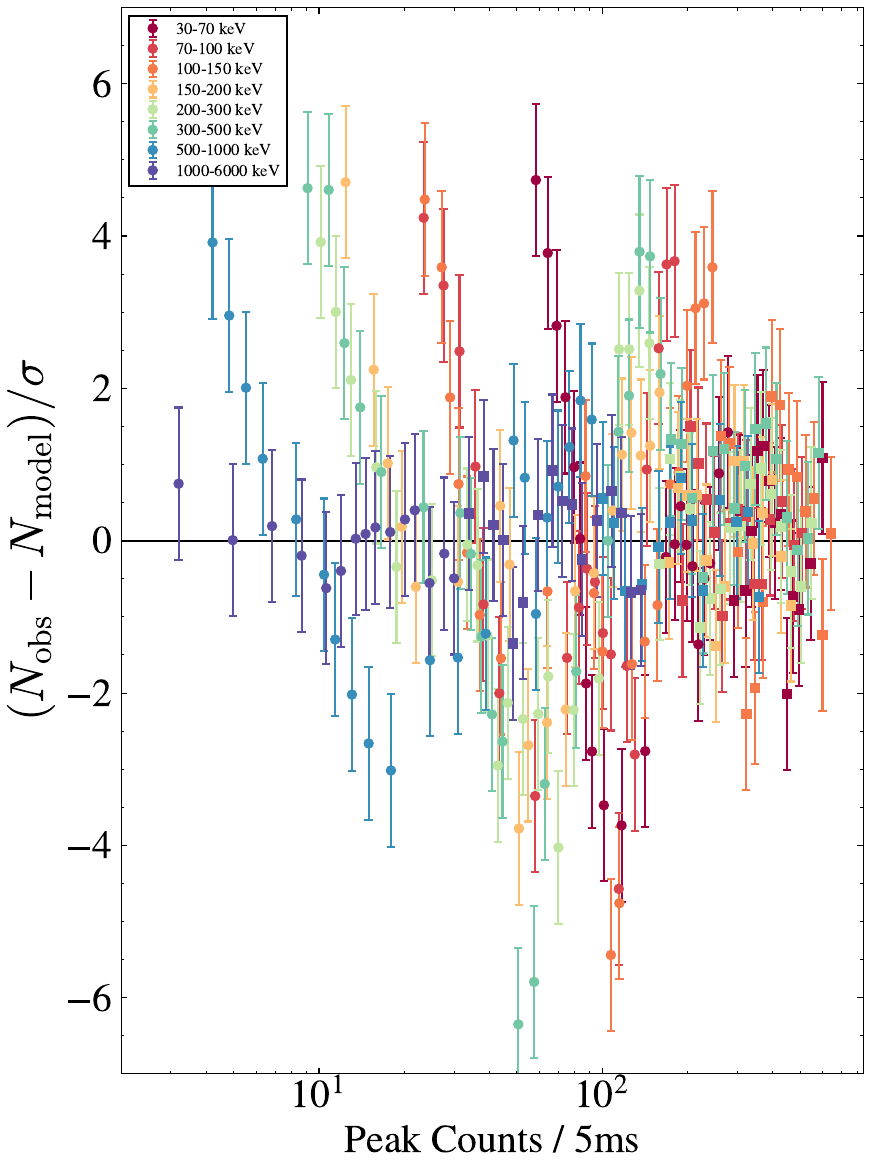}\put(-160, 225){\bf c}\\
\caption{\textbf{The residual for the fitting of waiting time ($\Delta t$), duration ($T$), and peak counts ($P$) in various energy bands.} 
\textbf{Panel (a)} The residual of $\Delta t$ of pulses with $1\sigma$ uncertainties in various energy bands are shown in different color. 
\textbf{Panel (b)} Same as panel (a), but for the $T$.
\textbf{Panel (c)} Same as panel (a), but for the $P$. }
\label{Residual}
\end{figure*}

\begin{figure*}[htp]
\centering
\includegraphics[width=\textwidth, angle=0]{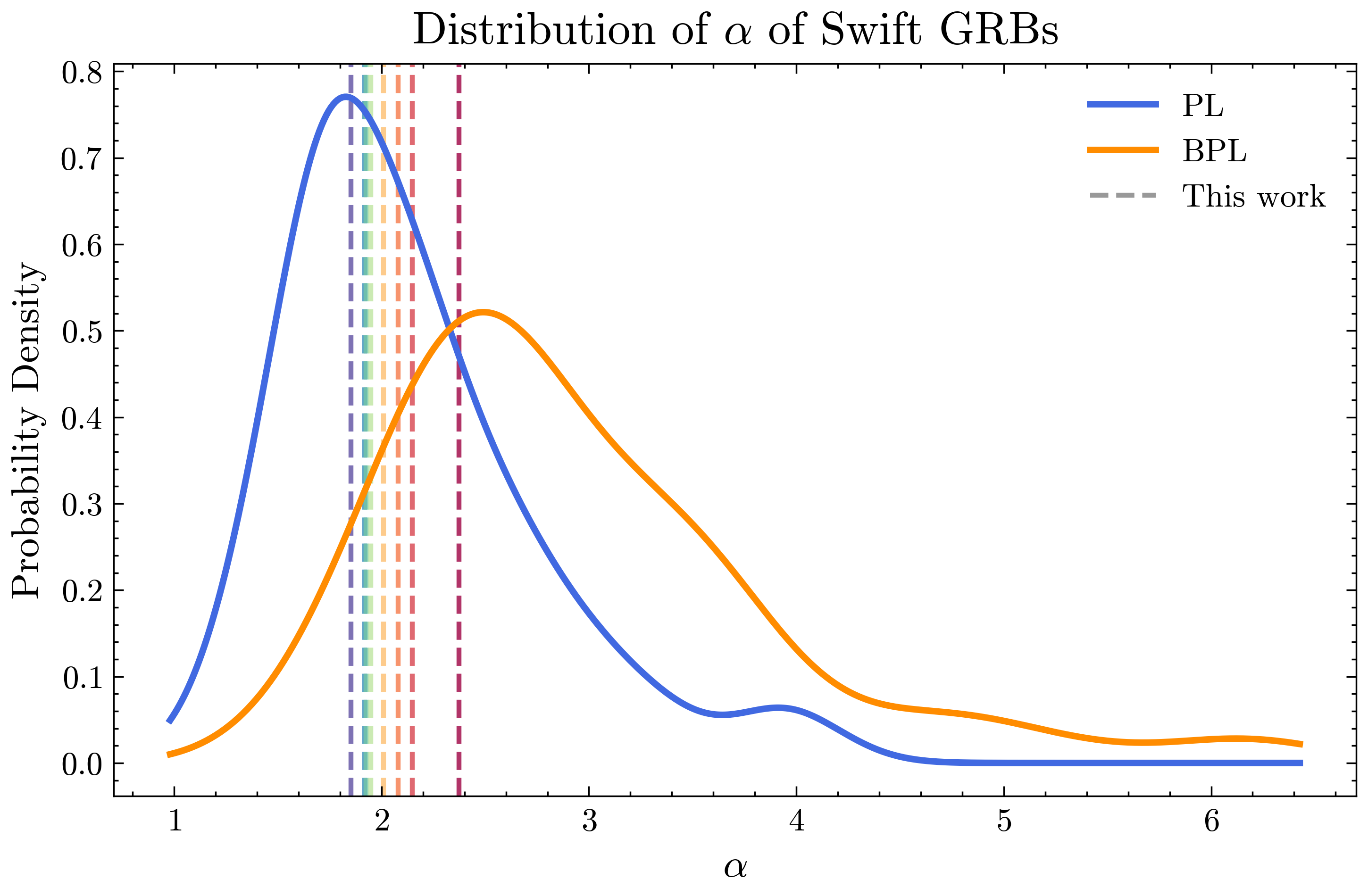}
\caption{\textbf{Comparison of PDS slopes with Swift GRBs.} 
 The PDS slopes of Swift GRBs, fitted by bent power-law (BPL) and power-law (PL) models, are collected from \cite{2016A&A...589A..98G}. The dashed lines show the fitted slopes obtained in this work, with colors from warm to cold corresponding to energy bands from low to high.}
\label{PDS_compare}
\end{figure*}

\begin{figure*}[htp]
\centering
\includegraphics[width=\textwidth, angle=0]{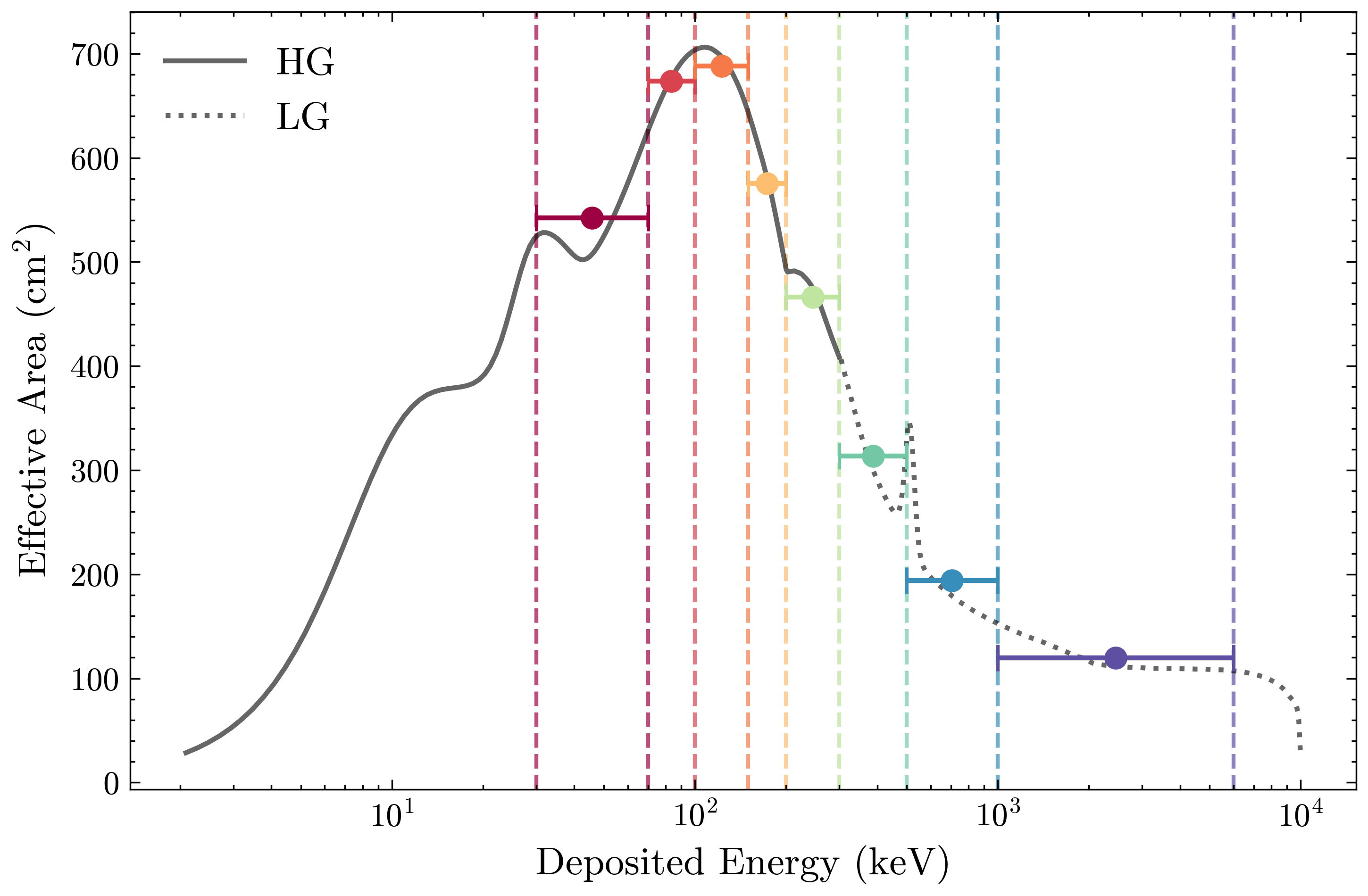}
\caption{\textbf{Total effective area of GECAM-B for GRB 230307A derived from the response matrix.} The vertical dashed lines and dots show the energy boundaries and the average effective area in each energy band, respectively, with colours from warm to cold indicating increasing energy.}
\label{eff_area}
\end{figure*}

\begin{figure*}[htp]
\centering
\includegraphics[width=\textwidth, angle=0]{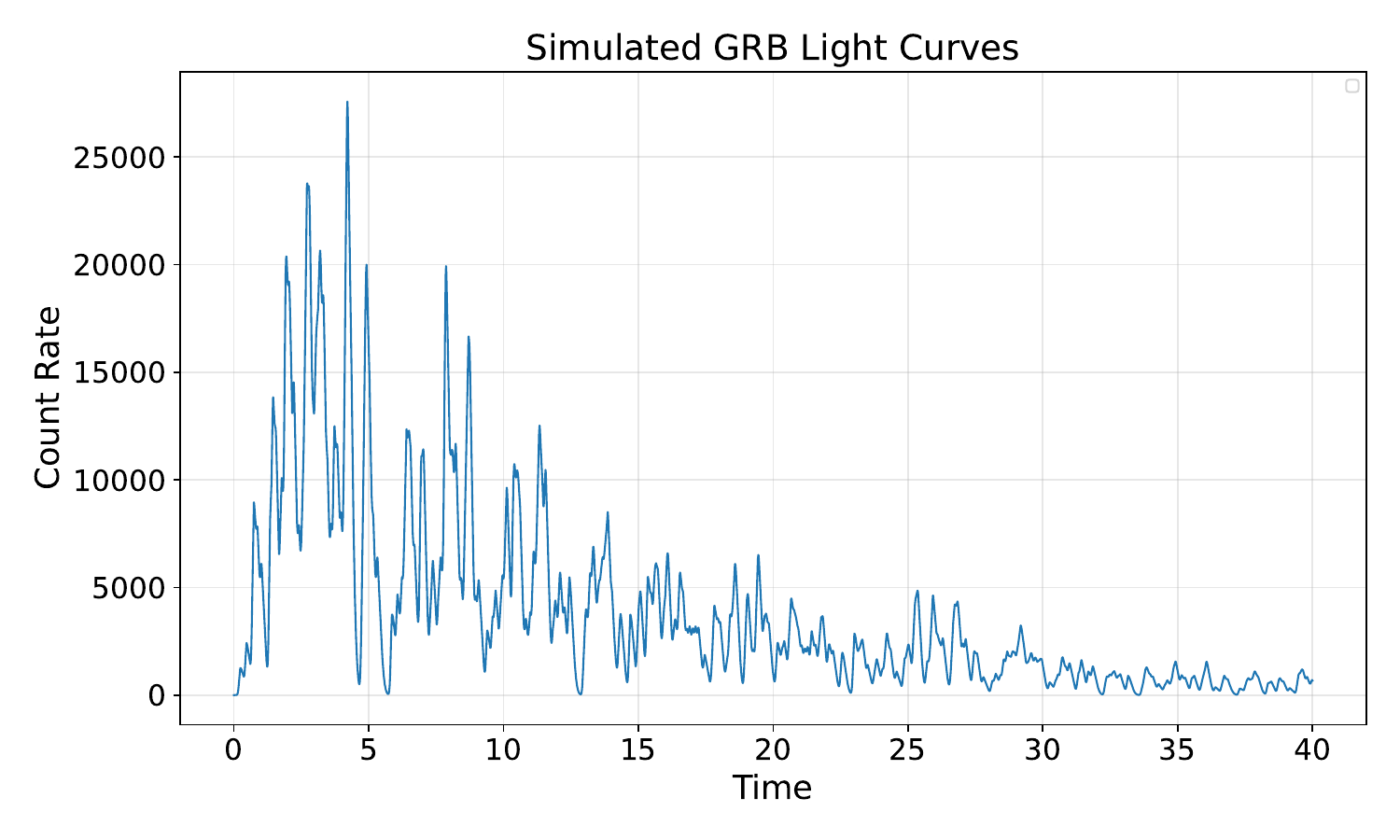}
\caption{\textbf{An example of the simulated light curve for the Poisson process.} The $x$ axis shows the time of 40 seconds and the $y$ axis shows the count rate.}
\label{example_simulation}
\end{figure*}

\begin{figure*}[htp]
\centering
\includegraphics[width=\textwidth, angle=0]{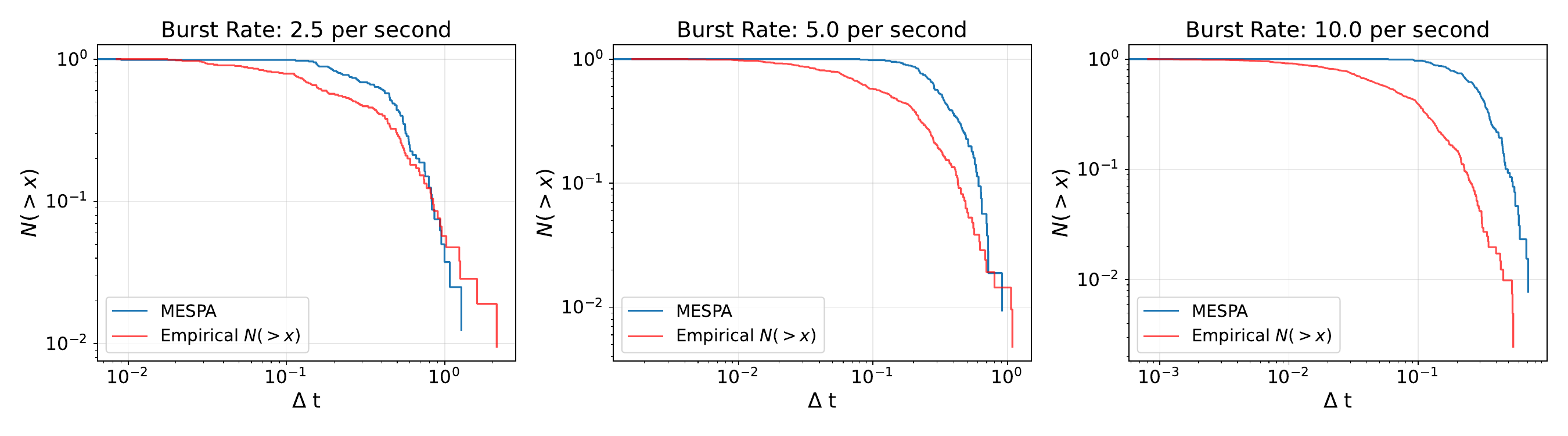}
\caption{\textbf{The waiting time distribution for the bursts simulated with Poisson process.} The three panels correspond to burst rates of 2.5, 5.0, and 10.0 burst per second, respectively. The red curves show the waiting-time distributions of all bursts generated by the Poisson process, whereas the blue curves show the results after burst selection using MESPA. As shown in the figure, the burst sample identified by MESPA tends to make the waiting-time distribution closer to an exponential distribution rather than a power-law form.}
\label{Poisson_simulation}
\end{figure*}

\begin{figure*}[htp]
\centering
\includegraphics[width=\textwidth, angle=0]{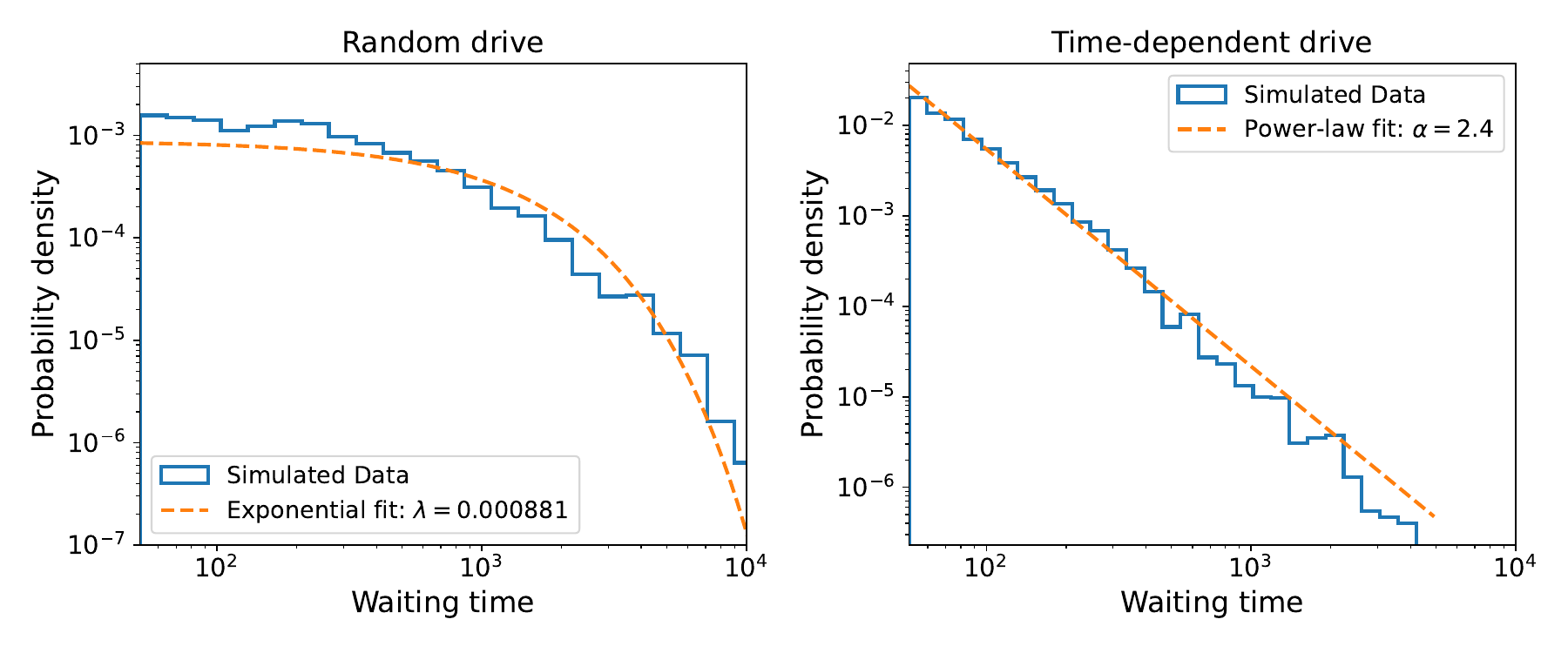}
\caption{\textbf{The waiting time distribution for the bursts generated by cellular automaton simulation.}The left panel shows the waiting-time distribution for random driving forces, which is well described by an exponential distribution. The right panel shows the waiting-time distribution for a driving force associated with a random-walk function, which follows a power-law form.}
\label{CA}
\end{figure*}

\end{document}